\documentclass[11pt]{article}

\usepackage[authoryear,round]{natbib}
\usepackage{amsmath,amssymb,amsfonts,amsthm,bbm,mathtools}
\usepackage{epic,eepic,epsfig,longtable}
\usepackage{multirow,verbatim}
\usepackage{array}
\usepackage{kotex}
\usepackage{todonotes}
\usepackage{epsfig}
\usepackage{setspace}
\usepackage{CJK}
\usepackage{color}
\usepackage[hyphens]{url}

\usepackage{rotating}
\usepackage{enumitem}

\usepackage[unicode=true]{hyperref}
\hypersetup{breaklinks=true,
            colorlinks=false,
            pdfborder={0 0 0}}

\makeatletter
\def\singlespace{\def\baselinestretch{1}\@normalsize}

\makeatletter
\def\singlespace{\def\baselinestretch{1}\@normalsize}

\numberwithin{equation}{section}

\renewcommand{\hat}{\widehat}

\newcommand{\bfm}[1]{\ensuremath{\mathbf{#1}}}

   \def\bA{\bfm A}

   \def\bD{\bfm D}  
\def\be{\bfm e}     
    
\def\bg{\bfm g}

     \def\RR{\mathbb{R}}

\def\bx{\bfm x}     
\def\by{\bfm y}

\def\calN{{\cal  N}}

\newcommand{\bfsym}[1]{\ensuremath{\boldsymbol{#1}}}

 \def\bfeta{\bfsym {\eta}}

 \def\btheta{\bfsym {\theta}}

\DeclareMathOperator{\E}{E}

\def\newpage{\vfill\eject}

\def\today{\ifcase\month\or
  January\or February\or March\or April\or May\or June\or
  July\or August\or September\or October\or November\or December\fi
  \space\number\day, \number\year}

\newdimen\biblioindent    \biblioindent=30pt

 at 8truept

\newcommand{\beq}{\begin{equation}}
  \newcommand{\eeq}{\end{equation}}
\newcommand{\beqn}{\begin{eqnarray}}
  \newcommand{\eeqn}{\end{eqnarray}}
\newcommand{\beqnn}{\begin{eqnarray*}}
  \newcommand{\eeqnn}{\end{eqnarray*}}

\long\def\comment#1{}

\allowdisplaybreaks
\usepackage{verbatim}
\renewcommand{\baselinestretch}{1.66}
\newtheorem{dfn}{Definition}
\newtheorem{thm}{Theorem}
\newtheorem{prop}{Proposition}

\newcounter{CondCounter}

\theoremstyle{definition}

\newtheorem{example}{Example}
\newtheorem{remark}{Remark}

\newcommand*\dif{\mathop{}\!\mathrm{d}}

\def \etadot   {\eta_{\boldsymbol{\cdot}}}

\begin{document}

\title{On a Notion of Graph Centrality Based on $L_1$ Data Depth} 

\author{
  Seungwoo Kang\footnote{kangsw0401@snu.ac.kr}~ and Hee-Seok Oh\footnote{Corresponding author: heeseok@stats.snu.ac.kr}\vspace{0.1in}\\
  Department of Statistics, Seoul National University\\
  Seoul 08826, Republic of Korea}
\date{\today}

\maketitle
\begin{abstract}
\noindent
A new measure to assess the centrality of vertices in an undirected and connected graph is proposed. The proposed measure, $L_1$ centrality, can adequately handle graphs with weights assigned to vertices and edges. The study provides tools for graphical and multiscale analysis based on the $L_1$ centrality. Specifically, the suggested analysis tools include the target plot, $L_1$ centrality-based neighborhood, local $L_1$ centrality, multiscale edge representation, and heterogeneity plot and index. Most importantly, our work is closely associated with the concept of data depth for multivariate data, which allows for a wide range of practical applications of the proposed measure. Throughout the paper, we demonstrate our tools with two interesting examples: the Marvel Cinematic Universe movie network and the bill cosponsorship network of the 21st National Assembly of South Korea. 
\end{abstract}

\noindent {\it Keywords}: Graph centrality; Data depth; Multiscale analysis; Visualization; Network data



\newpage


\section{Introduction} \label{sec:intro}

A fundamental and indispensable tool for analyzing graph data is a \emph{graph centrality} measure, which evaluates the prominence of vertices in a given graph structure \citep{sabidussi1966centrality,freeman1978centrality}. Several measures have been suggested based on different concepts of what constitutes a `central' vertex \citep[see, e.g.,][]{borgatti2006graph,rodrigues2019network}. The concept of centrality has been developed primarily in social network analysis and, more broadly, in the field of social sciences \citep{wasserman_faust_1994}. 

Meanwhile, in statistics, the concept of \emph{data depth} for multivariate data has been extensively studied since \cite{tukey1975mathematics}. Data depth is a multivariate generalization of a univariate rank, but in a center-outward manner, starting from the deepest point(s) and extending towards the outer points. Notable instances of data depth include half-space depth \citep{tukey1975mathematics}, simplical depth \citep{liu1990notion}, projection depth \citep{serfling2000general}, and $L_1$ data depth \citep{vardi2000multivariate}, among others. See \cite{serfling2000general,mosler2022choosing} for an extensive review. The data depth is a robust and nonparametric analytic tool for multivariate data. It is known for its powerful usage in robust estimation \citep{zuo2004stahel}, regression analysis \citep{rousseeuw1999regression}, and functional data analysis \citep{lopez2009concept,sun2011functional}.

Therefore, both graph centrality and data depth have a surprisingly comparable idea in that they measure the degree of `centralness' of a point or a vertex w.r.t.\ a given set of data. Nevertheless, there are a limited number of studies that establish a connection between these two fields of research. \cite{aamari2021graph} developed a data depth function by generating a neighborhood graph based on the provided multivariate data and calculating the centrality of that graph. \cite{cerdeira2021centrality} introduced a new centrality measure, termed Tukey centrality, which leverages the popular concept of half-space depth. However, as indicated in the latter study, the computation of Tukey centrality is NP-hard, so there is no reason to prefer it over the existing ones. To our knowledge, these are the only works connecting the two fields.

The aim of this paper is to introduce a new centrality measure, called $L_1$ centrality, for vertices of an undirected and connected graph, analogous to the $L_1$ data depth of \cite{vardi2000multivariate}. This measure can handle graphs with weights assigned to both vertices and edges in a straightforward manner. Nevertheless, the computation of this measure is simple and does not entail a higher computational cost than the existing centrality measures, unlike the approach proposed by \cite{cerdeira2021centrality}. Furthermore, due to the connection between our work and the notion of data depth for multivariate data, there are several practical possibilities for expanding the proposed measure based on the extensive literature on data depth. For instance, the $L_1$ centrality-based neighborhood and the local $L_1$ centrality introduced in this study are closely related to the concept of local depth in \cite{paindaveine2013depth}. In addition, we provide various graphical and multiscale analysis tools that rely on the $L_1$ centrality. The tools are based on an insightful interpretation of the suggested measure (Remark \ref{rmk:natural-order}), distinguishing our centrality measure from others. Through these tools, we demonstrate that even the basic concept of centrality has a wide range of practical uses. 

The rest of this paper is organized as follows. The $L_1$ centrality measure is defined in Section \ref{sec:L1cent}, and its properties are discussed, along with comparisons to the existing measures. Section \ref{sec:target-plot} presents a visualization tool, the target plot, that utilizes the $L_1$ centrality. Section \ref{sec:multiscale} expands upon the $L_1$ centrality measure using a multiscale approach. The $L_1$ centrality-based neighborhood of a vertex is precisely defined and utilized to derive the local $L_1$ centrality. The method for representing edges at multiple scales is also described. In Section \ref{sec:lorenz-curve}, we propose a group heterogeneity index to analyze graph data effectively at the group level. In Sections \ref{sec:L1cent}--\ref{sec:lorenz-curve}, the Marvel Cinematic Universe movie network is utilized to aid our explanation. Then, in Section \ref{sec:assembly}, we consider the bill cosponsorship network of 317 members in the 21st National Assembly of South Korea, which effectively demonstrates the practicality of the tools developed in the study. Finally, Section \ref{sec:discussion} provides concluding remarks with further discussion. All technical proofs are in the Appendix. 

All methods and data sets used in this paper are provided via the R package \textbf{L1centrality} \citep{l1cent}, available from the Comprehensive R Archive Network. Also, the codes for reproducing the figures and analysis in this paper are available from \url{https://github.com/seungwoo-stat/L1centrality-paper}.


\section{$L_1$ Centrality Measure} \label{sec:L1cent}
\subsection{Notations and Review of Graph Centrality Measures} \label{subsec:background}

Denote a graph by $G = (V, E)$, where $V = \{v_1,\dots,v_n\}$ is a set of vertices (nodes, points, or actors), and $E$ is a set of edges (ties, or lines) connecting pairs of distinct vertices. The number of vertices, $n = |V|$ ($|\cdot|$ indicates the cardinality of the set), is often referred to as a \emph{graph size}. When edges do not have direction, i.e., every connection from vertex $i$ to $j$ has a connection from $j$ to $i$, we call the graph \emph{undirected}. Otherwise, the graph is directed. Hereafter, we assume that the given graph is undirected. 

The adjacency matrix is denoted by $\bA = (A_{ij})_{n\times n}$, where $A_{ij}=1$ if $v_i$ and $v_j$ are directly connected by an edge and 0 otherwise. A sequence of edges connecting a set of vertices is called a \emph{path}. A graph is said to be \emph{connected} when it is possible to reach any vertex from any other vertex, that is, when a path always exists between any two vertices. Based on the paths, define $d(v_i, v_j)$, the \emph{geodesic distance} between $v_i$ and $v_j$, as the shortest path (\emph{geodesic path}) length between $v_i$ and $v_j$. Here, the path length is the sum of the weights of the edges of that path. When the edges of a graph all have weight 1, i.e., (edge) unweighted graph, the path length is simply equal to the number of edges in that path. Of course, this distance function satisfies all the properties of the usual distance function, such as triangle inequality and symmetry.

As noted, edges can have positive weights, which denote the distance between two nodes through the edge. Likewise, vertices can also have nonnegative weights, indicating the importance of that node. This could refer to the contextual information that reflects the significance of the vertices. For notational simplicity, we call vertex weights \emph{multiplicities} and edge weights just \emph{weights} since a weighted graph usually refers to a graph with edge weights, not vertex multiplicities. 

The notion of centrality quantifies the `centralness' of each vertex in a given graph structure. However, there is no consent to this concept, which has led to several definitions of centrality. We focus on three intuitive graph centrality measures listed below \citep{freeman1978centrality}. In the following definitions, we only consider connected and unweighted (all edge weights and all vertex multiplicities are set to 1) graphs.

\begin{enumerate}[noitemsep]
\item The \emph{degree centrality} regards a vertex with many direct connections to other vertices as central. Specifically, degree centrality at vertex $v_i$ is defined as $\sum_{j=1}^n A_{ij}$.
\item The \emph{closeness centrality} defines a vertex as central if it is close to all other vertices. In the formula, it is defined as the reciprocal of the sum of the distances to other nodes. That is, closeness centrality at vertex $v_i$ is $1/\sum_{j\neq i}d(v_i, v_j)$. 
\item The \emph{betweenness centrality} perceives that the vertex in the `middle' of the geodesic paths has more influence. For $v_i$, it is quantified as $\sum_{j<k,j\neq i\neq k}g_{jk}(v_i)/g_{jk}$, where $g_{jk}$ denotes the number of geodesic paths connecting $v_j$ and $v_k$, and $g_{jk}(v_i)$ indicates the number of geodesic paths between $v_j$ and $v_k$ that pass through $v_i$. 
\end{enumerate}
Other centrality measures include eigenvector centrality \citep{bonacich1972factoring}, hubs and authorities \citep{kleinberg1999authoritative}, 
and much more variants. See e.g., \citet[Chapter 5]{wasserman_faust_1994}, \cite{borgatti2006graph}, \cite{rodrigues2019network} and references therein. 

Each notion of graph centrality can possibly be generalized for graphs with weights and multiplicities. For example, denoting the multiplicity of vertex $v_j$ as $\eta_j >0$, the degree centrality of $v_i$ can be generalized as $\sum_{j\neq i}\eta_j A_{ij}/d(v_i,v_j)$. However, this generalization is forced, and there is no suitable explanation for why this should be the generalized form of the degree centrality, meaning that any variation similar to the above can be proposed. Likewise, the closeness centrality and betweenness centrality can be generalized in an arbitrary way. This is because these centrality measures were originally defined for unweighted graphs without consideration of handling weights. To our knowledge, there is no consensus on generalizing these centrality measures to (vertex and edge, but especially vertex) weighted graphs. 


\subsection{Motivation and Definition} \label{subsec:l1cent}

Although there are many definitions of centrality, there is a similarity. It first defines a measure and then perceives the vertex with the highest measure as the most central node (hereafter, `center'). For example, for the betweenness centrality, one may ask, `What is the center vertex based on the betweenness centrality?' One would answer, of course, `The vertex with the highest betweenness centrality.' 

Our approach goes in the opposite direction from these traditional approaches. First, we define the notion of a center vertex. Next, we describe a centrality measure from this notion of center vertex. Roughly speaking, our approach quantifies how much a given node is related to the center vertex. This is the essential feature of the proposed measure, which we elaborate on after defining it. 

Now, assume that a given graph has edge weights and vertex multiplicities. However, the graph is still assumed to be undirected and connected. For disconnected graphs, the proposed centrality measure can be applied to each connected component (maximal connected subgraph). The treatment of directed graphs will be discussed in Section \ref{sec:discussion}. To this end, the notion of the center vertex, \emph{graph median}, is defined below. 

\begin{dfn}[Graph median \citep{hakimi1964optimum}] \label{def:g-median}
Given an undirected, connected graph $G=(V,E)$ with $V=\{v_1,\dots,v_n\}$ and multiplicity $\eta_1,\dots,\eta_n \geq 0$ for each vertex, $v_i$ is called a \emph{graph median} if it minimizes $\sum_{j=1}^n \eta_j d(v_i,v_j)$.
\end{dfn}

The definition of graph median can be seen as an analogous concept of a multivariate median or the $L_1$ median \citep{small1990survey,vardi2000multivariate}. However, it is important to note that the $L_1$ median is unique unless multivariate points are collinear, i.e., all lie on a line \citep{small1990survey}, whereas the graph median can have more than one element. For example, all vertices of a complete graph (without weights or multiplicities) are the graph median. Denote the set of graph medians of graph $G$ with multiplicities as $m(G; \eta_1,\dots,\eta_n)$. Based on the notion of graph median, the $L_1$ centrality measure is defined. 

\begin{dfn}[$L_1$ centrality measure]
Suppose that $G=(V,E)$ is an undirected, connected graph with $V=\{v_1,\dots,v_n\}$ and multiplicity $\eta_1,\dots,\eta_n\geq 0$ for each vertex. Assuming that $\etadot \coloneqq \sum_{j=1}^n \eta_j>0$, the \emph{$L_1$ centrality} of vertex $v_k$ is defined as
\begin{align*}
C(v_k)\coloneqq 1 - \inf\left\{w\geq 0: v_k \in m\left(G; \frac{\eta_1}{\etadot},\dots,\frac{\eta_k}{\etadot} + w,\dots,\frac{\eta_n}{\etadot}\right)\right\}.
\end{align*}
\end{dfn}

In other words, the $L_1$ centrality of a node is one minus the minimum amount of multiplicity that must be incremented for that node to become a graph median. This definition leads to the following closed-form formula for calculating the $L_1$ centrality measure, 
\begin{align}
C(v_k) &= 1 - \inf\left\{w\geq 0: \sum_{i=1}^n \frac{\eta_i}{\etadot} d(v_k,v_i) = \min_{j=1,\dots,n}\left(\sum_{i=1}^n \frac{\eta_i}{\etadot} d(v_j,v_i) + w d(v_j,v_k)\right) \right\} \nonumber\\
&= 1-  \max_{j\neq k} \left\{\frac{\sum_{i=1}^n \eta_i \{d(v_k,v_i) - d(v_j,v_i)\}}{\etadot d(v_j,v_k)}\right\}^+, \label{eq:l1-cent}
\end{align}
where $\{\cdot\}^+ \coloneqq \max\{0,\cdot\}$. 

We make three notes. Firstly, the graph median always has an $L_1$ centrality of 1. Secondly, the $L_1$ centrality is always between 0 and 1 due to the triangle inequality applied to equation \eqref{eq:l1-cent}. Hence, a node with an $L_1$ centrality close to one indicates a central vertex. Moreover, since the measure is independent of the graph size $n$, i.e., the measure is normalized, it is possible to compare the $L_1$ centrality values of vertices in different graphs. Notice that the above three centrality measures depend on the graph size and require suitable normalizations \citep{freeman1978centrality}. Lastly, when $\eta_i = 0$, the $i$th term in the numerator of equation \eqref{eq:l1-cent} is eliminated. In other words, assigning a multiplicity of zero to a particular vertex means disregarding its influence while calculating the $L_1$ centrality. Therefore, in most cases, all multiplicities are assigned a positive value unless one intends to disregard that vertex purposely.

\begin{remark}\label{rmk:natural-order}
The essence of the proposed approach is that the centrality measure is defined based on the notion of graph median, which means that the proposed centrality ranks the relative importance of vertices w.r.t.\ the center vertex. Specifically, by incrementing the multiplicity of $v_k$ by $1-C(v_k)$, $v_k$ becomes the graph median. Hence, $1-C(v_k)$ is the amount of multiplicity required to replace the original graph median by vertex $v_k$. In other words, the greater the $L_1$ centrality, the higher the relevance of the vertex to the graph median. However, this is not the case with the above three centrality measures. For example, a vertex with a high betweenness centrality does not imply that the vertex is relevant to the vertex with the highest betweenness centrality. This concept is a useful and important aspect of the $L_1$ centrality, enabling further analysis tools in Sections \ref{sec:target-plot} and \ref{sec:multiscale}.
\end{remark}

\begin{example}[$L_1$ centrality of the Marvel Cinematic Universe movies]\label{ex:MCU-l1cent}
To facilitate the explanation of the proposed centrality measure, we proceed on a small toy network of Marvel Cinematic Universe (MCU) movies. This data set was first used by \cite{choi2021heavy} to analyze graphs with signals. The data set consists of 32 movies from the MCU released between 2008 and 2023. Each movie represents one vertex in the graph. Since these movies share a common universe or plots, they often share casts. When there is at least one common cast between movies $i$ and $j$, the edge is connected,  with weight $|A_i\cup A_j|/|A_i\cap A_j|$. Here $A_i$ is a set of cast from movie $i$. In other words, if the proportion of common casts is large, the path length for that edge is small. We also used the worldwide gross of each movie (in USD, archived from IMDb  (\url{https://www.imdb.com}) on Nov.\ 3rd, 2023) as the multiplicity of each vertex. Hence, the network is an undirected and connected graph with 32 vertices (with positive multiplicities) and 278 edges (with weights). 

\begin{figure}
\center
\includegraphics[width=\textwidth]{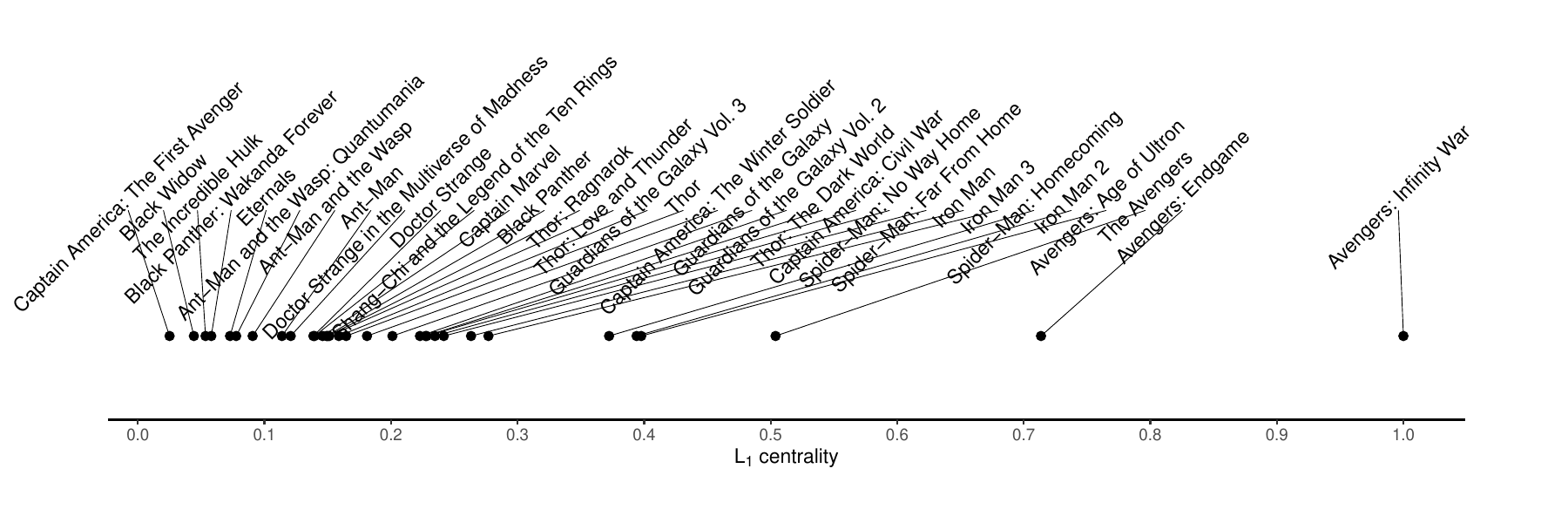}
\vspace{-3em}
\caption{$L_1$ centralities of each vertex (movie) in the MCU movie network.}\label{fig:mcu-l1cent}
\end{figure}

The $L_1$ centrality of each movie is shown in Figure \ref{fig:mcu-l1cent}. As expected, four \emph{Avengers} series (\emph{Avengers: Infinity War} (the graph median), \emph{Avengers: Endgame}, \emph{The Avengers}, \emph{Avengers: Age of Ultron}), in which many heroes overlap with the other series, were the top four movies with the highest $L_1$ centrality. Furthermore, it is worth mentioning that these movies had high worldwide gross, highlighting their significant positions within the network. In other words, the context surrounding the box office performance of each movie is taken into account while calculating the $L_1$ centrality. The subsequent section mathematically addresses the influence of the multiplicity of a particular vertex on its $L_1$ centrality.
\end{example}

Before closing this section, we connect the proposed $L_1$ centrality to data depth literature. The definition of the $L_1$ centrality is analogous to the $L_1$ data depth of \cite{vardi2000multivariate}. However, the $L_1$ data depth is defined for any point (not necessarily a data point) in the sample space, while the $L_1$ centrality is only defined for vertices. Nevertheless, the notion of the $L_1$ centrality can be extended to an arbitrary point on existing edges. To simplify the definition, in Definition \ref{def:g-median}, we intentionally restricted the graph median to be found at the vertices. However, \cite{hakimi1964optimum} called the graph median an \emph{absolute median} and proposed a more generalized concept---it can be any point on the graph, not necessarily a vertex, and can be any point on an existing edge. Based on this definition, we may define $L_1$ centrality measure for an arbitrary point on the existing edges. In this paper, however, we only stick to vertex centrality. Graphs that an arbitrary point on the edge has an interpretable meaning are limited, and we leave this topic for future research.

\subsection{Theoretical Properties}


\begin{thm}[Properties of $L_1$ centrality] \label{thm:property}
Suppose that $G=(V,E)$ is an undirected, connected graph with $V=\{v_1,\dots,v_n\}$, multiplicities $\eta_1,\dots,\eta_n\geq 0$ for each vertex, and $\etadot \coloneqq \sum_{j=1}^n\eta_j >0$. The $L_1$ centrality measure has the following properties:
\begin{itemize}[noitemsep]
\item[\emph{(P1)}] \emph{Scale invariance}: It is invariant to (positive) multiplicative transformations of vertex multiplicities and edge weights. 
\item[\emph{(P2)}] \emph{Maximality}: It is maximized to 1 if and only if the given vertex is the graph median. If $\eta_i/\etadot \geq 1/2$, then $C(v_i)=1$. If $\eta_i/\etadot > 1/2$, $v_i$ is the unique vertex with $C(v_i)=1$. 
\item[\emph{(P3)}] \emph{Minumum value}: $C(v_i)\geq \min\{2\eta_i/\etadot, 1\}$.
\item[\emph{(P4)}] \emph{Minimum at infinity}: Suppose that the subgraph induced by deleting vertex $v_1$ is connected. When $v_1$ is moved to infinity, that is, $d(v_1,w)\to\infty$ for all $w\in V\setminus \{v_1\}$, and $d(v_1,w)/d(v_1,w')\to 1$ for all $w,w'\in V\setminus\{v_1\}$, then $C(v_1)\to\min\{2\eta_1/\etadot,1\}$.
\end{itemize}
\end{thm}

We assert that properties (P2) and (P3) align with our intuitive understanding of what constitutes a central vertex. The greater the multiplicity, the more likely the vertex is to have a large $L_1$ centrality. Throughout this paper, Theorem \ref{thm:property} will be used several times as we develop analysis tools based on the $L_1$ centrality. 

In connecting the notion of graph centrality to the data depth studies, it is worth referring to the concept of statistical depth. In \cite{serfling2000general}, four properties of the data depth function were proposed, and a data depth function satisfying all these properties is called \emph{statistical depth}. They are (i) affine invariance, (ii) maximality at the center, (iii) monotonicity relative to the deepest point, and (iv) vanishing at infinity. Property (P1) can be viewed as an analog of (i), and (P2) as (ii), and (P4) as (iv). However, we could not not find a property analogous to (iii). 

\subsection{Comparison and Computation}

We compare the $L_1$ centrality measure to the three centrality measures outlined in Section \ref{subsec:background}. The $L_1$ centrality is most relevant to the closeness centrality of the three measures. If the closeness centrality is generalized as $(\sum_{i=1}^n \eta_i d(v_i,v_k))^{-1}$ for $v_k$, the $L_1$ centrality and the closeness centrality identify the same set of the center vertex (the graph median). Hence, the proposed measure and closeness centrality share a consensus on what the most central vertex is. However, they differ in the way they quantify and rank the centrality of vertices other than the graph median. We assert that the $L_1$ centrality is a more sophisticated measure than the closeness centrality. Recall equation \eqref{eq:l1-cent}, 
\begin{align*}
C(v_k) &= 1-  \max_{j\neq k} \left\{\frac{\sum_{i=1}^n \eta_i \{d(v_k,v_i) - d(v_j,v_i)\}}{\etadot d(v_j,v_k)}\right\}^+.
\end{align*}
In the numerator, we see that the reciprocals of the $v_k$'s closeness centrality appear. The larger the closeness centrality of $v_k$, the larger $C(v_k)$ is likely to be. However, what is also considered in the above equation is the closeness centrality of nearby vertices. Suppose that we have two vertices with the same closeness centrality. One is at the terminal, and the other is surrounded by many other vertices. A neighbor vertex connected by the former terminal vertex has a much higher closeness centrality because all geodesic paths from the terminal vertex must pass through the neighbor vertex; thus, the former terminal vertex is likely to have a small $L_1$ centrality. 

In addition, as mentioned in Remark \ref{rmk:natural-order}, the $L_1$ centrality is a natural ordering of vertices w.r.t.\ the graph median. This is what makes our measure different from the other three measures. It allows us to visualize the network based on the center, define a natural notion of the neighborhood that takes into account the whole graph structure, and localize the proposed centrality measure, enabling multiscale analysis. 

\begin{example}[Centrality measures applied to the MCU movie network]
We compare the degree, betweenness, and closeness centralities to the $L_1$ centrality measure. To this end, we used the MCU movie network with the same positive multiplicity for all vertices because the three centrality measures cannot be easily generalized for graphs with multiplicities. However, we kept edge weights since the closeness and betweenness centralities can incorporate edge weights while the degree centrality cannot. 

The Pearson correlation coefficients (Spearman rank correlation coefficients) of the $L_1$ centrality with the degree, betweenness, and closeness centrality measures were 0.6604 (0.7400), 0.8778 (0.6037), and 0.7317 (0.7852), respectively. Thus, the proposed $L_1$ centrality was positively correlated with the existing measures but was not equivalent to any of the three measures. That is, the rankings induced by the centrality measures are different. Another thing to note is that in the MCU movie network, the betweenness centrality treats half of the vertices as centrality zero, making comparing these vertices difficult. On the other hand, the $L_1$ centrality never treats any $L_1$ centrality as zero (Theorem \ref{thm:property} (P3)), and there are no ties between the $L_1$ centralities of 32 vertices. Scatter plots comparing the $L_1$ centrality with the three measures are provided in the Appendix \ref{app:compare}.
\end{example}

We next explain the practical computation procedure of the $L_1$ centrality. Suppose that the geodesic distance matrix $\bD = (d(v_i,v_j))_{n\times n}$ is computed. Then, the computation of the $L_1$ centrality, $C(v_k)$, for all $k=1,\dots,n$ can be completed within ${\mathcal{O}}(n^2)$ computations (in big-O notation). Specifically, the vector of $L_1$ centralities, $C(V)\coloneqq(C(v_1),\dots,C(v_n))^\top$ can be computed using the following formula, 
\begin{align}\label{eq:l1cent-mat}
C(V) = \mathbf{1}_n - \frac{1}{\etadot} \left\{\texttt{rowmax}\left(\frac{\bD\bfeta\mathbf{1}_n^\top - \mathbf{1}_n\bfeta^\top\bD}{\bD}\right)\right\}^+,
\end{align}
where $\mathbf{1}_n \coloneqq (1,\dots,1)^\top$ and $\bfeta\coloneqq(\eta_1,\dots,\eta_n)^\top$. Moreover, division by a matrix is defined element-wise, and $\{\cdot\}^+$ is applied to each element of the vector. The \texttt{rowmax} operation indicates the maximum element in each row, and when taking the \texttt{rowmax} operation, we ignore the diagonal terms or define $0/0 = 0$.

Based on the above argument, the computational cost of the $L_1$ centrality is not greater than that of the closeness centrality and the betweenness centrality. This is because the bottleneck in computing the $L_1$ centrality is not in the computation of equation \eqref{eq:l1cent-mat} but in the computation of the geodesic distance matrix $\bD$. The computation of the geodesic matrix for an (edge) weighted graph can be achieved using the approach suggested by \cite{dijkstra1959note} or the Floyd--Warshall algorithm \citep{floyd1962algorithm}. The time complexity of these algorithms is ${\mathcal{O}}(n^3)$. The time complexity can be reduced for sparse graphs using a special data structure \citep{Fredman1984fibo}, but it is not lower than ${\mathcal{O}}(n^2)$. Therefore, the computation of any centrality measure that needs to construct the geodesic distance (e.g., the closeness centrality and betweenness centrality) possesses a time complexity greater than or equal to the time complexity of the $L_1$ centrality computation. 


\section{Target Plot} \label{sec:target-plot}

In this section, we propose a novel graph visualization tool that aims to embed a graph onto a two-dimensional plane in a way that effectively represents the structural information of each vertex, specifically the $L_1$ centrality. As in the previous section, we examine a graph with $n$ vertices that is undirected, connected, and has weighted edges. However, we impose a constraint that all vertices have the same multiplicity, meaning that we exclusively examine vertex unweighted graphs. The reason for this restriction will be explained shortly. 

Various graph drawing approaches are available for representing graphs on a two-dimensional plane, and each approach has a distinct perspective on what constitutes a visually pleasing representation. The algorithms proposed by \cite{kamada1989algorithm} and \cite{fruchterman1991graph} are often used in this domain. However, to our knowledge, none of the existing algorithms take graph centrality into account when deciding where to place points in their plot. Typical methods do not place the most central vertex in the center of the plot, nor do they put the least central vertex in the periphery. 


In light of Remark \ref{rmk:natural-order}, it is intuitive to position the graph median at the central point and the remaining vertices distributed around concentric circles whose radii correspond to their $L_1$ centrality. A graph is visually represented by the plot in a way that effectively conveys the structural characteristics of each vertex’s centrality. In pursuit of this objective, we present a method of plotting graphs called \emph{target plot}.    

\begin{enumerate}[noitemsep]
\item We aim to determine a configuration of $n$ points on a two-dimensional plane, with each vertex represented as an individual point. Let $\bx_i\in \RR^2$ denote the point corresponding to vertex $v_i$ ($i=1,\dots,n$). In circular coordinates, this point is denoted as $(r_i,\theta_i)\in [0,\infty)\times [0,2\pi)$. The target plot constraints $\bx_i$ to lie on a concentric circle with radii $r_i = -\log C(v_i)$, so the graph median is placed at the center of the circles. The log transformation converts $L_1$ centralities in the interval $(0,1]$ to a radius in the interval $[0,\infty)$. This transformation also reflects the observation that a vertex distant from all other vertices is likely to have a low $L_1$ centrality, according to equation \eqref{eq:l1-cent} and Theorem \ref{thm:property} (P4). However, this observation may not be valid if there are distinct multiplicities on the vertices, i.e., a vertex far from all other vertices may exhibit high $L_1$ centrality if it has a high degree of multiplicity due to Theorem \ref{thm:property} (P3). For this reason, in this section, we focus only on unweighted vertex graphs. 

\item Next, we use the nonmetric multidimensional scaling method \citep[nonmetric MDS;][]{kruskal1964multidimensional,kruskal1964nonmetric} with the constraint specified in 1. In essence, the goal is to find a configuration that retains the geodesic distances on the graph as much as possible while the constraint is enforced. Specifically, we aim to find a configuration of points that minimizes a metric called \emph{stress}, adopted from \cite{kruskal1964multidimensional}:
\begin{align*}
S &\coloneqq \sqrt{\frac{S^*}{T^*}},\quad
\text{where}\quad S^* \coloneqq \sum_{i,j} (d_{ij} - \hat{d}_{ij})^2, \quad T^*\coloneqq \sum_{i,j}d_{ij}^2.
\end{align*}
Here, $d_{ij}\coloneqq \|\bx_i-\bx_j\|$, where $\|\cdot\|$ denotes the usual Euclidean norm, and $\hat{d}_{ij}$ are chosen to minimize the stress metric given $d_{ij}$, subject to the following monotonicity condition:
\begin{align*}
\begin{cases}
\text{whenever } d(v_i, v_j) < d(v_k, v_l), &\text{then } \hat{d}_{ij} \leq \hat{d}_{kl},\\
\text{whenever } d(v_i, v_j) = d(v_k, v_l), &\text{then } \hat{d}_{ij} = \hat{d}_{kl}.
\end{cases}
\end{align*}

There is a fast and efficient algorithm for fitting $\hat{d}_{ij}$. See \cite{kruskal1964nonmetric} for details.

A starting configuration is set using the classical MDS \citep{mardia1978some}. If $v_i$ is the graph median, $\bx_i=\mathbf{0}$; otherwise, $\bx_i$ is set to $r_i (\by_i - \by_*)/\|\by_i - \by_*\|$, where $\by_i\in\RR^2$ is the point representing $v_i$ as a result of classical MDS, and $v_*$ is the graph median with $\by_*$ being the corresponding point derived from classical MDS. Given the configuration, it can be readily shown that
\begin{align*}
\frac{\partial S}{\partial\theta_i} 
&= \sqrt{\frac{T^*}{S^*}}\frac{1}{T^*}\sum_{j\neq i}r_ir_j\sin(\theta_i-\theta_j)\left(1-\frac{S^*}{T^*}-\frac{\hat{d}_{ij}}{d_{ij}}\right).
\end{align*}
The gradient descent method is employed to iteratively update the value of $\theta_i$ until convergence. Specifically, denoting $\btheta \coloneqq (\theta_1,\dots,\theta_n)^\top$ and $\bg \coloneqq \partial S/\partial\btheta = (\partial S/\partial\theta_1,\ldots,\partial S/\partial\theta_n)^\top$, the gradient descent method is applied in a similar way to  \cite{kruskal1964nonmetric}:
$
\btheta_\text{new} = \btheta_\text{old} - \alpha \frac{\bg}{\operatorname{mag}(\bg)},
$
where $\operatorname{mag}(\bg) \coloneqq \|\bg\|/\sqrt{\sum_{i=1}^n r_i^2}$ (the relative magnitude of $\bg$), and $\alpha$ is the step size with an initial value of 0.2 and is set to a smaller value by $\alpha_\text{new} = 0.95\cdot\alpha_\text{old}$ in each iteration. 
The above process is repeated until $\operatorname{mag}(\bg)$ is small enough.

\item From 1 and 2, we obtain a configuration $(r_i, \theta_i)$ representing the vertex $v_i$. After plotting each point on a two-dimensional plane, contours of concentric circles are drawn to assist in recognizing and understanding the graph structure. For example, in Figure \ref{fig:mcu-target} (b), the radii of the four concentric circles correspond to the first (25\%) to fourth (maximum) quartiles of the $L_1$ centralities. Hence, a quarter of the points lie within the smallest circle, and these points correspond to the vertices with the highest 25\% $L_1$ centrality. From its appearance, it is clear why we call the resulting figure a target plot. This plot can be enhanced by including the color or shape of the points to facilitate more appropriate interpretation. For instance, there might be identifiable patterns in which the least central vertices share, indicating that it is desirable to remove vertices that exhibit those patterns before further analysis. This analysis will be demonstrated in Section \ref{sec:assembly}. 
\end{enumerate}

\begin{example}[Target plot of the MCU movie network]
Figure \ref{fig:mcu-target} shows two distinct methods of graph representation on a two-dimensional plane. Panel (a) displays the plot proposed by \cite{fruchterman1991graph}, and panel (b) shows the proposed target plot. Note that we used the same multiplicity for all vertices in this example. The target plot represents the structural information of $L_1$ centrality, showing the most central vertex (black point) and the least central vertex. In contrast, panel (a) does not convey this information. 

\begin{figure}
\center
\includegraphics[width=\textwidth]{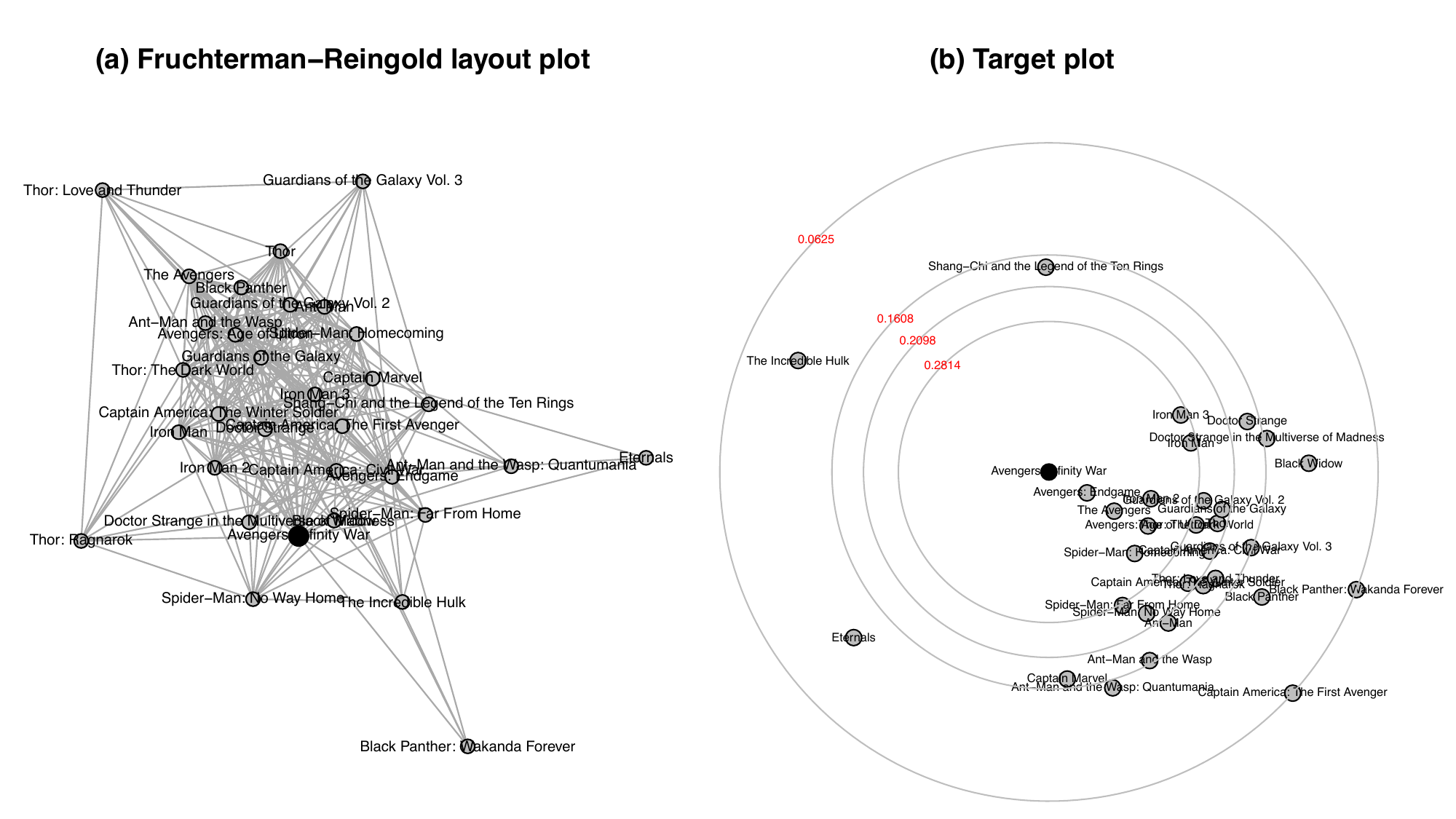}
\vspace{-3mm}
\caption{(a) \cite{fruchterman1991graph} plot of the MCU movie network. (b) Target plot of the MCU movie network. The four concentric circles represent the quartiles of $L_1$ centrality, with the values indicated in red. In both figures, the graph median (\emph{Avengers: Infinity War}) is denoted by a black circle, while gray circles represent the remaining vertices.}\label{fig:mcu-target}
\end{figure} 

In the target plot, three points are far away from the rest: \emph{The Incredible Hulk}, \emph{Shang-Chi and the Legend of the Ten Rings}, and \emph{Eternals}. These three movies are (the only) outliers in this distribution when computing the average geodesic distance of each movie to the others. In other words, the average distances of these three movies exceed the sum of 1.5 times the interquartile range and the third quartile of average distances. As a result, the target plot accurately represents the inherent structure of the network.
\end{example}

Before closing this section, we make three remarks. First, the concept of the target plot is not applicable to other graph centrality measures. A small problem is that the radii of each point cannot be determined using a straightforward log transformation. A crucial aspect is connected to Remark \ref{rmk:natural-order}. For example, plotting vertices with high-degree centralities near the center of the concentric circles is utterly inadequate since it conveys the misleading impression that vertices with high-degree centralities are related. Second, the gradient descent method can converge to a local minimum of the stress measure, not an overall minimum. However, as \cite{kruskal1964nonmetric} pointed out, this is not a significant problem. The local minimum configuration is not the final result of the analysis, and it should be of interest if it is meaningful. Refer to the discussion in \citet[Section 5]{kruskal1964nonmetric}. 
Finally, the target plot resembles 
the \emph{depth contour}, a visualization tool commonly used in data depth literature \citep[e.g.,][]{zuo2000structural}. Depth contours, similar to our concentric circles, aid in identifying points with high and low data depths. 

\section{Local $L_1$ Centrality Measure} \label{sec:multiscale}

The degree centrality is a  \emph{local} measure considering only the vertices directly connected to a particular vertex. On the other hand, the closeness and betweenness centralities are \emph{global} measures of centrality because they consider all vertices in the graph while calculating these measures. The $L_1$ centrality, in the same sense, is a global centrality measure. However, the structural properties of vertices can vary at both local and global scales, and all graph centrality measures, including the $L_1$ centrality, can only capture one of the two. 

This section introduces a local extension of $L_1$ centrality based on the \emph{local depth} proposed by \cite{paindaveine2013depth}, where the degree of locality can be selected to a range of levels. This extension provides a \emph{multiscale} view of a single graph with the centrality values at various locality levels. As described below, the local measure is derived by conditioning the graph to suitable $L_1$ centrality-based neighborhoods. 

\subsection{$L_1$ Centrality-Based Neighborhood} \label{subsec:cent-nb}

Referring back to Remark \ref{rmk:natural-order}, vertices with high $L_1$ centrality can be considered neighborhoods of the graph median. This approach inherently considers the multiplicities of vertices and the geodesic distances when determining neighborhoods w.r.t.\ the graph median. 

\begin{figure}
\center
\includegraphics[width=.7\textwidth]{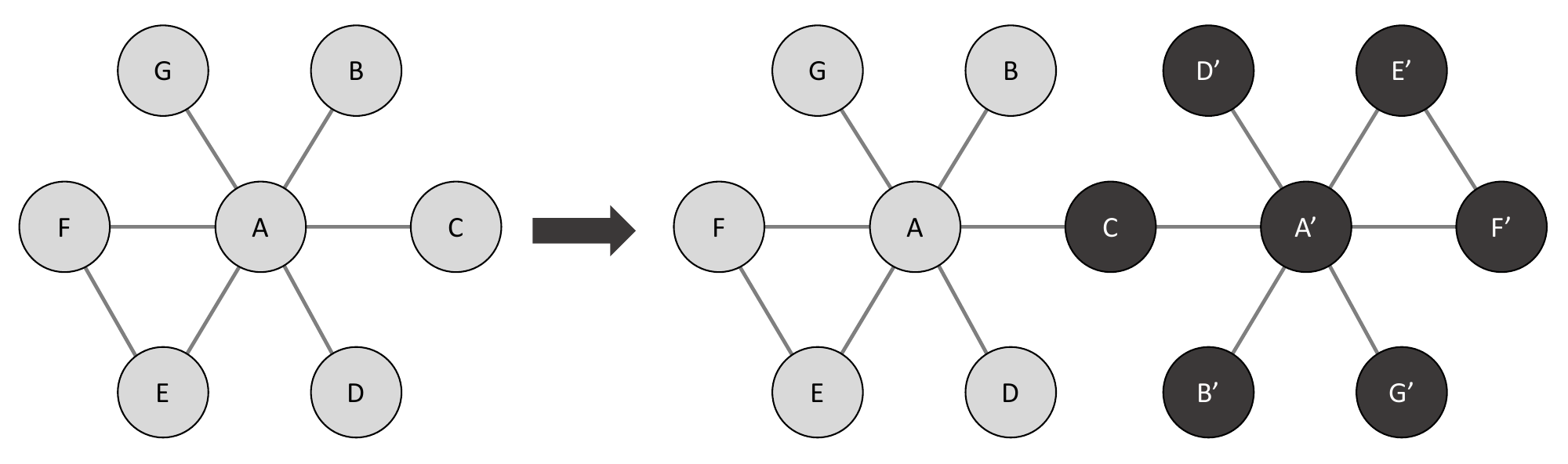}
\vspace{-3mm}
\caption{A given graph (left) and symmetrization of the given graph w.r.t.\ vertex C (right).}\label{fig:d-symm}
\end{figure}

However, this approach is not applicable to vertices other than the graph median. To this end, we perform a symmetrization procedure analogous to \cite{paindaveine2013depth} that symmetrized a multivariate distribution for a specific point to derive depth-based neighborhoods of that point. Similarly, we symmetrize the graph w.r.t.\ a particular vertex, say $v_i$, and set position $v_i$ to the graph median. The process is depicted in Figure \ref{fig:d-symm}, which shows a graph with seven vertices in the left panel. To create a symmetric graph for vertex C, we duplicate the entire graph, including the vertex multiplicities and edge weights, and overlap the original vertex C with the duplicated vertex C. Therefore, the multiplicity of vertex C is doubled, and the other copied vertices have the same multiplicity as the original, e.g., $\eta_\text{A} = \eta_\text{A'}$, where $\eta_\text{A}$ stands for the multiplicity of vertex A. This symmetrization procedure makes vertex C a graph median based on the following proposition.  

\begin{prop}\label{prop:d-symm}
Suppose that $G=(V,E)$ is an undirected, connected graph with $V = \{v_1,\dots,v_n\}$, a nonnegative multiplicity $\eta_j$ for vertex $v_j$, $j=1,\ldots,n$, and $\sum_{j=1}^n \eta_j > 0$. If $G$ is symmetrized w.r.t.\ vertex $v_i\in V$, then $v_i$ is a graph median. If $\eta_i > 0$, $v_i$ is the unique graph median in the symmetrized graph w.r.t.\ $v_i$.
\end{prop}

Therefore, the $L_1$ centralities of vertices in the symmetrized graph w.r.t.\ vertex $v_i$ will serve as a natural ordering of vertices in relation to $v_i$. This results in the establishment of the $L_1$ centrality-based neighborhood. 

\begin{dfn}[$L_1$ centrality-based neighborhood]
The order $\alpha\in[0,1]$ \emph{$L_1$ centrality-based neighborhood} of vertex $v_i$ is the set of vertices in the original graph, with $L_1$ centrality in the symmetrized graph w.r.t.\ $v_i$ larger than or equal to the $100(1-\alpha)$\% quantile of these centralities. 
\end{dfn}

Of the $2n-1$ vertices in the symmetrized graph, each copied vertex has the same $L_1$ centrality as the original vertex. Therefore, of the $n$ vertices in the original graph, about $n\alpha$ vertices are in the $L_1$ centrality-based neighborhood of each vertex. The parameter $\alpha$ acts as a locality parameter, with smaller values corresponding to smaller neighborhoods. However, it would be demanding to construct a new geodesic matrix of size $(2n-1)\times (2n-1)$ for the $L_1$ centrality computation each time a graph is symmetrized w.r.t.\ a specific vertex. Fortunately, it is unnecessary to compute a new distance matrix; this can be done by modifying only the multiplicities within the original graph. 

\begin{prop}\label{prop:new-distance-mat}
Suppose that $G=(V,E)$ is an undirected, connected graph with $V=\{v_1,\dots,v_n\}$ and a nonnegative multiplicity $\eta_j$ for vertex $v_j$, $j=1,\ldots,n$. If $G$ is symmetrized w.r.t.\ vertex $v_i$, the $L_1$ centrality of vertex $v_k$ in the resulting graph is equal to the $L_1$ centrality of vertex $v_k$ in the original graph, but with the multiplicity of vertex $v_i$ substituted by $\etadot + \eta_i$, where $\etadot = \sum_{j=1}^n \eta_j > 0$. 
\end{prop}

For example, to calculate the $L_1$ centrality of vertex A in the right panel of Figure \ref{fig:d-symm}, it is enough to calculate the $L_1$ centrality of vertex A in the left panel, where the multiplicity of vertex C is substituted with $\etadot + \eta_\text{C}$ (here, $\etadot = \eta_\text{A} + \dots + \eta_\text{G}$). The geodesic distance matrix is calculated only once for the original graph and is subsequently utilized for computing $L_1$ centrality in the symmetrized graph. 

Another significant implication of Proposition \ref{prop:new-distance-mat} is that the $L_1$ centrality-based neighborhood strikes a balance between vertices with high $L_1$ centrality (in the original graph) and vertices near the vertex of interest. Since the symmetrization w.r.t.\ vertex $v_i$ makes the graph median $v_i$, vertices near $v_i$ are likely to have high $L_1$ centrality in the symmetrized graph. From another perspective, the $L_1$ centrality of the symmetrized graph is the same as the $L_1$ centrality of the original graph with multiplicities modified. Therefore, vertices with high $L_1$ centrality in the original graph are likely to have high $L_1$ centrality in the symmetrized graph. 


\begin{example}[$L_1$ centrality-based neighborhood of \emph{Spider-Man: No Way Home}] 
By utilizing the MCU movie network and assigning vertex multiplicities based on the worldwide gross, we calculate the $L_1$ centrality of the graph symmetrized w.r.t.\ the vertex \emph{Spider-Man: No Way Home}. The corresponding values are shown in Figure \ref{fig:mcu-local-spiderman}. The $L_1$ centrality-based neighborhood of the vertex \emph{Spider-Man: No Way Home} is obtained using this ranking. For example, the order 5/32 $L_1$ centrality-based neighborhood refers to the five movies with the highest $L_1$ centralities in Figure \ref{fig:mcu-local-spiderman}. These vertices have large $L_1$ centralities in the original graph (\emph{Avengers} series; see Figure \ref{fig:mcu-l1cent}) or are strongly linked to the vertex of interest (\emph{Spider-Man: Homecoming} and \emph{Spider-Man: Far From Home} are the two nearest (in geodesic distance) neighbors of \emph{Spider-Man: No Way Home} in the MCU movie network). 
\begin{figure}
\center
\includegraphics[width=\textwidth]{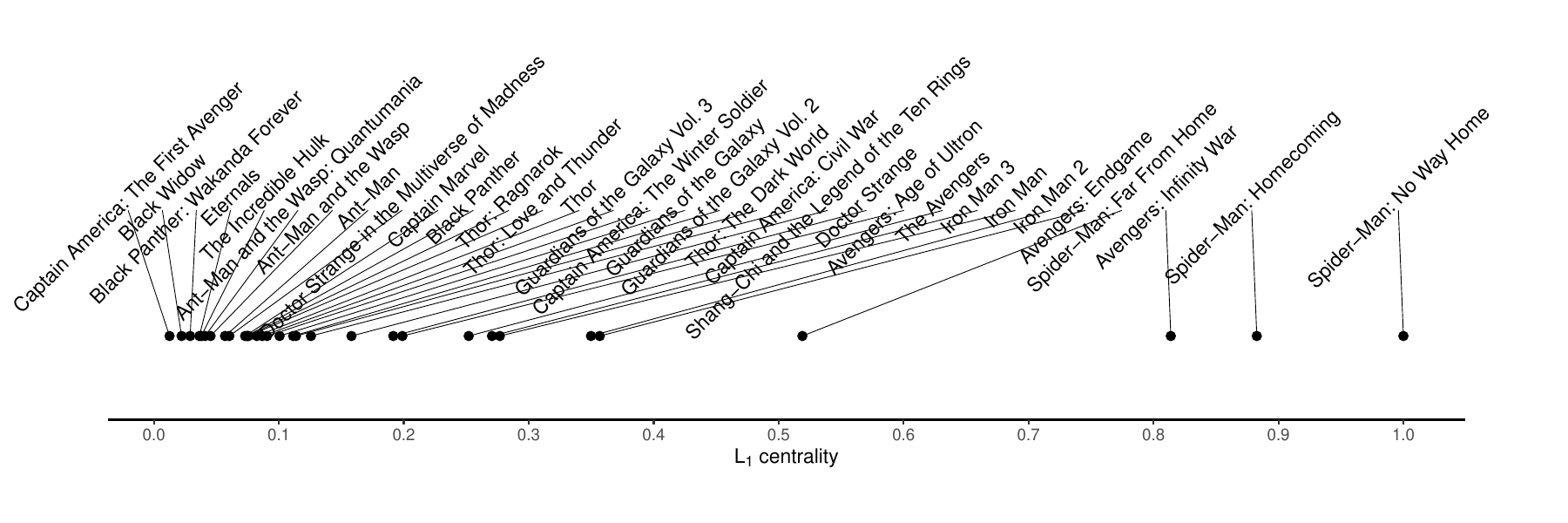}
\vspace{-10mm}
\caption{$L_1$ centralities of each vertex (movie) in the MCU movie network symmetrized w.r.t.\ \emph{Spider-Man: No Way Home}.}\label{fig:mcu-local-spiderman}
\end{figure}

Although the $L_1$ centrality-based neighborhood is developed to define the local $L_1$ centrality in the next section, it can also serve as a valuable tool on its own. For example, it can be used to recommend pertinent films to users who are interested in \emph{Spider-Man: No Way Home} while simultaneously considering the significance (high $L_1$ centrality) in the entire network and relevance (small geodesic distance) to the target movie. 
\end{example}

\subsection{Local $L_1$ Centrality} \label{subsec:local-cent}

Defining \emph{local $L_1$ centrality} based on the above $L_1$ centrality-based neighborhood is straightforward: for each vertex, the graph is conditioned on the $L_1$ centrality-based neighborhood, and the local $L_1$ centrality is then computed from that graph. More precisely, we condition the graph in the sense defined below. 

\begin{dfn}[Local $L_1$ centrality]\label{def:locall1}
The order $\alpha$ \emph{local $L_1$ centrality} of vertex $v_k$, denoted as $C_\alpha(v_k)$, is the $L_1$ centrality of $v_k$ in the original graph, but only considering the $L_1$ centrality-based neighborhood of $v_k$ up to order $\alpha$ during the computation. Specifically, denoting the set of order $\alpha$ $L_1$ centrality-based neighborhood of $v_k$ as $\calN_\alpha(v_k)$,
\begin{align}
C_\alpha(v_k) &= 1-  \max_{j: v_j\in\calN_\alpha(v_k)\setminus\{v_k\}} \left\{\frac{\sum_{i: v_i\in\calN_\alpha(v_k)} \eta_i \{d(v_k,v_i) - d(v_j,v_i)\}}{d(v_j,v_k) \sum_{i:v_i\in\calN_\alpha(v_k)}\eta_i}\right\}^+. \label{eq:local-l1-cent}
\end{align}
\end{dfn}

In contrast to equation \eqref{eq:l1-cent}, the range of summation and maximum operations are confined to the $L_1$ centrality-based neighborhood. Therefore, calculating the local $L_1$ centrality $C_\alpha(v_k)$ by equation \eqref{eq:local-l1-cent} is equivalent to computing the $L_1$ centrality of $v_k$ using equation \eqref{eq:l1cent-mat}, where the distance matrix and the multiplicity vector are replaced by a submatrix and a subvector of the original. The submatrix is formed by choosing the rows and columns corresponding to the indices of the order $\alpha$ $L_1$ centrality-based neighborhood, and the subvector is a subset of the multiplicity vector corresponding to the order $\alpha$ $L_1$ centrality-based neighborhood. Hence, the properties of the $L_1$ centrality, including Theorem \ref{thm:property}, also apply to the local $L_1$ centrality. Clearly, the order 1 local $L_1$ centrality is identical to the $L_1$ centrality defined in Section \ref{sec:L1cent}. This measure will be referred to as \emph{global $L_1$ centrality}. 

Alternative to Definition \ref{def:locall1}, one might use the \emph{subgraph} of the original graph, which is created by only considering the order $\alpha$ $L_1$ centrality-based neighborhood of $v_i$ and edges that connect these vertices. However, we prefer Definition \ref{def:locall1} for two reasons. First, when utilizing the subgraph, it is necessary to recalculate the distance matrix, which requires more computational effort, in contrast to the definition above, which only uses the submatrix of the original distance matrix. Second, especially for small values of $\alpha$, it cannot be ensured that the resulting subgraph is connected. 

The definition of the local $L_1$ centrality is not to be confused with the $L_1$ centrality in the symmetrized graph defined in Section \ref{subsec:cent-nb}. The symmetrization procedure is used to identify the $L_1$ centrality-based neighborhood. Once the neighborhood is identified, the $L_1$ centrality values in the symmetrized graph are not used. The local $L_1$ centrality is computed in the original graph conditioned on these neighboring vertices. 

Concerning Theorem \ref{thm:property} (P3), the local $L_1$ centrality is generally higher than the global $L_1$ centrality as the lower bound ($2\eta_i/\etadot$) increases. Therefore, rather than focusing on the absolute value of the local $L_1$ centrality, we propose to look at its relative value, i.e., its rank, compared to the local $L_1$ centrality of other vertices of the same order. Given these points, the local $L_1$ centrality provides a direct method to investigate the structure of a graph at different levels, depending on the value of $\alpha$. In Section \ref{sec:assembly}, we show how to leverage the local $L_1$ centrality to perform a multiscale analysis of a graph. 

\subsection{Multiscale Edge Representation} \label{subsec:edge-rep}

The local $L_1$ centrality defined in the previous section does not rank vertices based on the graph median. Instead, it quantifies the relevance of each vertex in relation to its \emph{local median}, which is the graph median of the conditioned graph. Naturally, adopting the target plot in Section \ref{sec:target-plot} for the local $L_1$ centralities with $\alpha <1$ is not suitable since the concept of ranking w.r.t.\ the graph median no longer applies to local $L_1$ centralities. However, in this section, we present a multiscale visualization tool that uses the local $L_1$ centrality. The objective of this visualization method is to represent a given graph with approximately $n=|V|$ directed edges at different locality levels. The fundamental idea is similar to that of the target plot. 

At a global level, the graph median can be regarded as the central point of the graph, as in the target plot. A given graph can be represented as a directed graph, where each vertex has an edge pointing to the graph median. At a locality level $\alpha$, we can construct a directed graph where each vertex is connected to its local median vertex determined at locality level $\alpha$. Edges are connected to each local median if more than one local median exists. Therefore, for every value of $\alpha$, the graph is represented as a directed graph containing about $n$ edges. The visualization of the directed graph also makes it easy to identify the local medians, facilitating further examination of that vertex. 

\begin{figure}
\center
\includegraphics[width=\textwidth]{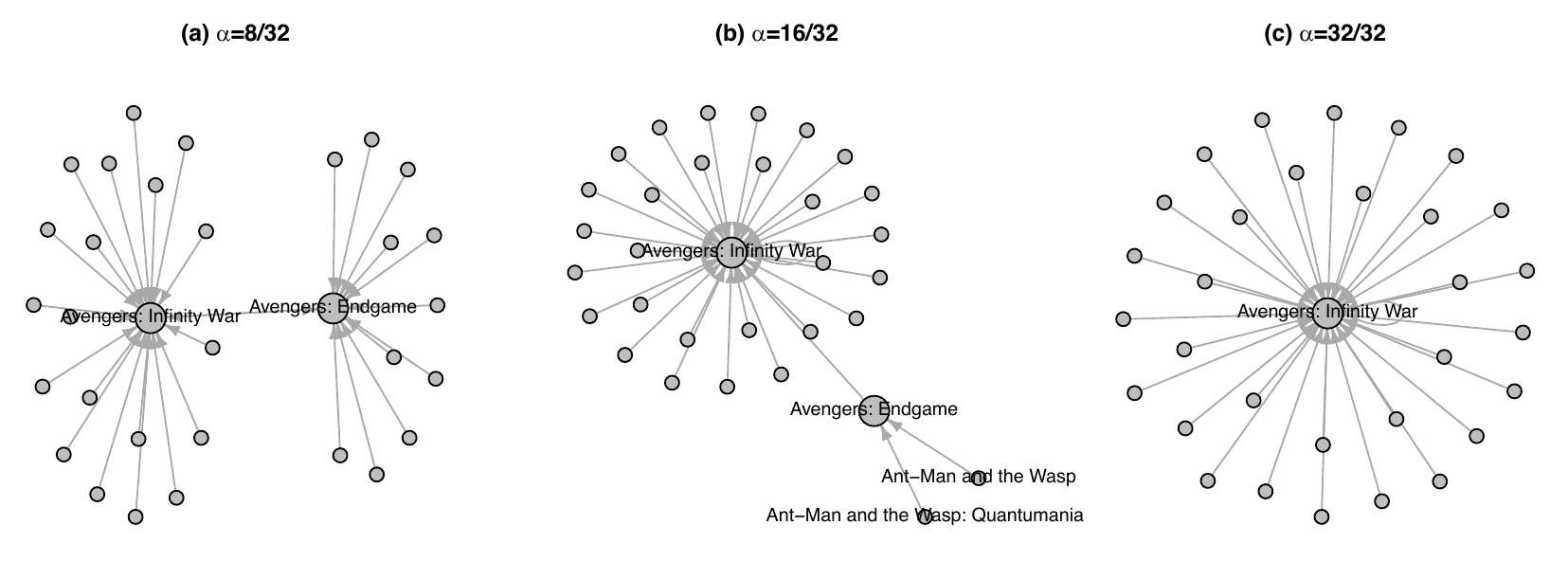}
\vspace{-10mm}
\caption{Edge representation of the MCU movie network at different levels of $\alpha$. Larger vertices represent the local medians.}\label{fig:mcu-edge}
\end{figure}

\begin{example}[Multiscale edge representation of the MCU movie network] 
Figure \ref{fig:mcu-edge} shows the directed graphs of the MCU movie network with multiplicities set to the worldwide gross. We selected three locality levels: $\alpha = 8/32, 16/32, 32/32$, which means that we consider approximately 8, 16, and 32 vertices for each vertex when conditioning the graph and determining the $L_1$ centrality-based neighborhood and local median. In panel (a), it is evident that each vertex has either \emph{Avengers: Infinity War} or \emph{Avengers: Endgame} as a local median. The same applies to panel (b), where only two vertices from the \emph{Ant-Man and the Wasp} series connect to \emph{Avengers: Endgame} as a local median. Most vertices designate \emph{Avengers: Infinity War} as the local median, which also serves as the graph median. Panel (c) provides a graph representation at the global level, where all vertices have edges directed toward the graph median. When comparing Figure \ref{fig:mcu-target} (a) to this multiscale edge representation, it is clear that the latter is significantly more comprehensible. The multiscale edge representation allows for a more precise visualization of the graph's global and local structures, providing a multiscale perspective of a given graph.
\end{example}

\section{Group Heterogeneity Index} \label{sec:lorenz-curve}

This section provides a visualization tool and an index representing heterogeneity in the proposed centrality measures for groups of vertices or the entire graph. The proposed visualization tool and index are important for many types of analysis. For example, it can be used to evaluate if a group or graph exhibits any peculiar structural characteristics, detect trends across several groups or graphs (e.g., graphs that change over time), or explore the relationship of the heterogeneity of the group’s vertices to other attributes of the group. Furthermore, it can be employed to compare various groups and graphs and categorize them into similar types \citep{badham2013commentary}.

Several indices have been developed to quantify the heterogeneity of centrality measures. Some of the prominent measures in this field are the concept of \emph{centralization} by \cite{freeman1978centrality}, the \emph{variance} index of \cite{snijders1981degree}, and the notion of \emph{hierarchization} by \citet[Chapter 14]{coleman1964introduction}. Recently, \cite{badham2013commentary} discussed using the Gini coefficient to measure heterogeneity; the author argued that it has favorable theoretical characteristics compared to the previously discussed indices. 

However, the Gini coefficient has yet to be well studied as a descriptive measure of centrality heterogeneity in network science. Interestingly, the Lorenz curve and Gini coefficient have been studied in data depth. \cite{liu1999multivariate} used the Lorenz curve and Gini coefficient to establish a metric for quantifying kurtosis in multivariate data. Given that the notion of graph centrality in this study is based on a specific data depth, the purpose of this section is to revisit the Lorenz curve and Gini coefficient as a tool for analyzing the heterogeneity of the proposed global and local $L_1$ centrality measures. Here is a detailed description of the benefits of using these tools for global and local $L_1$ centralities. 

The Lorenz curve is defined as follows. Given a univariate cumulative distribution function (cdf) $F$ with a nonnegative support, it is a plot of $(p, L(p))$, with
\begin{align}
L(p) \coloneqq \frac{\int_{0}^p F^{-1}(t) \dif t}{\int_{-\infty}^\infty s \dif F(s)} = p\times \frac{\E(X\mid X\leq F^{-1}(p))}{\E(X)}, \quad 0\leq p\leq 1, \label{eq:lorenz}
\end{align}
where $F^{-1}(t) \coloneqq \inf\{v\in\RR: F(v)\geq t\}$ and $X\sim F$ \citep{gastwirth1971general}. By definition, $L(p)$ is nondecreasing in $p$ and $L(p)\leq p$. Furthermore, $L(1) = 1$ and $L(0)=0$. Given $m$ centrality measurements, $C_\alpha(v_1),\dots,C_\alpha(v_m)\geq 0$, $F$ of \eqref{eq:lorenz} is substituted by the empirical cdf $F_m(v) = \frac{1}{m}\sum_{i=1}^m I(C_\alpha(v_i)\leq v)$, where $I$ denotes the indicator function. The Lorenz curve of the centrality measurements serves as an intuitive visual indicator of heterogeneity. The greater the deviation of the curve from the diagonal $L(p)=p$, the more it indicates that the group of $m$ vertices has a diverse structural attribute in a particular graph, i.e., their $L_1$ centralities show a high level of heterogeneity. 

The Gini coefficient, defined as twice the area between the Lorenz curve and the diagonal $L(p) = p$, is a simple and widely used metric for quantifying the level of diversity in income and wealth, making it valuable for comparing multiple distributions. When $F$ is replaced with $F_m$ using the measurements $C_\alpha(v_1),\dots,C_\alpha(v_m)$, it can be readily verified that the Gini coefficient $G$ can be represented as  
\begin{align*}
G = \frac{\sum_{i=1}^m \sum_{j=1}^m |C_\alpha(v_i) - C_\alpha(v_j)|}{2m^2 \bar{C}_\alpha},
\end{align*}
where $\bar{C}_\alpha \coloneqq (1/m)\sum_{i=1}^m C_\alpha(v_i)$. It is the expected difference of the centrality measurements with a suitable normalization. This also shows that the Gini coefficient is an appropriate index for quantifying heterogeneity among centrality values. Consistent with the results of the Lorenz curve, the larger the heterogeneity index, the less similar the nodes are in terms of centrality. 

There are several advantages to using the Lorenz curve and Gini coefficient to quantify the heterogeneity of the proposed $L_1$ centrality. First, the Lorenz curve provides a visual diagnostic tool for comparing several groups or graphs rather than directly summarizing centrality values as a single number. Second, these tools are scale-invariant. As mentioned, local $L_1$ centralities generally exhibit higher values than global $L_1$ centralities. Consequently, the proposed method can be applied to compare heterogeneity across multiple locality levels, even if the scales of $L_1$ centralities across several locality levels differ. Third, the Gini coefficient is inherently standardized to 0 and 1. Some indices, such as centralization and variance, require a normalization procedure that normalizes the index w.r.t.\ the highest feasible value for a graph with the same number of vertices  \citep{freeman1978centrality,snijders1981degree,coleman1964introduction}. Moreover, these procedures only apply to graphs without weights, and it remains uncertain how similar normalization procedures may be implemented for the edge-weighted graphs considered in this study. Thus, the normalization procedure limits the applicability of given indices to the entire graph. In contrast, the Lorenz curve and Gini coefficient can be utilized for a set of vertices, not necessarily the entire graph, with the index inherently standardized.


\begin{example}[Heterogeneity index of the MCU movie network]
We generate the Lorenz curve for the global $L_1$ centrality distribution of the MCU movie network using several multiplicities: (i) the same value, (ii) the worldwide gross, and (iii) the reciprocal of the worldwide gross. Figure \ref{fig:mcu-lorenz} shows that the multiplicity of worldwide gross places the Lorenz curve below the curve with equivalent multiplicities ($G=0.3339$). This indicates that the $L_1$ centrality exhibits a higher level of heterogeneity, i.e., a larger Gini coefficient ($G=0.4085$), when measured with the worldwide gross as a multiplicity. That is, the centrality of the central nodes increases while the centrality of the outliers decreases. A simple regression analysis using worldwide gross as the independent variable and the $L_1$ centrality (with equal multiplicity) as the dependent variable confirms this view. The regression coefficient for the worldwide gross is statistically significant ($p$-value $\leq 5\times 10^{-6}$).  

\begin{figure}
\center
\includegraphics[width=.6\textwidth]{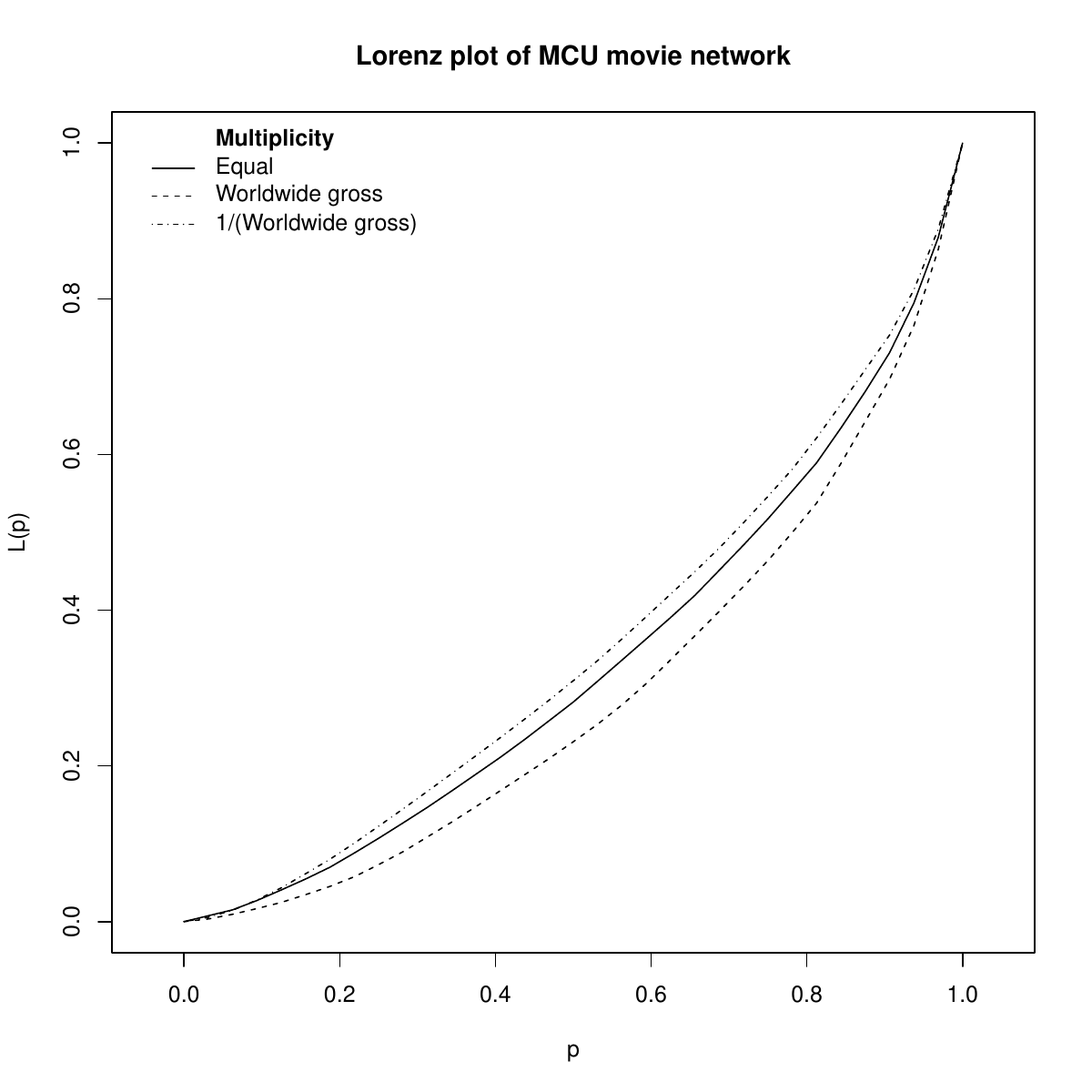}
\vspace{-5mm}
\caption{Lorenz curve of the MCU movie network with multiplicities set to various values.} \label{fig:mcu-lorenz}
\end{figure}

Alternatively, by defining the multiplicity as the inverse of the worldwide gross, we observe an upward shift of the Lorenz curve, indicating a decrease in the heterogeneity index ($G=0.2943$). This example shows that using the Lorenz curve and Gini coefficient as an index of heterogeneity is consistent with our understanding that assigning greater importance (multiplicity) to the central nodes amplifies the heterogeneity of the resulting centrality. 
\end{example}


\section{Application: South Korea's National Assembly Bill Cosponsorship Network} \label{sec:assembly}

In this section, we focus on the network of legislators in the 21st National Assembly of South Korea (May 30th, 2020--May 29th, 2024). Our main emphasis is on analyzing the network at multiple scales, demonstrating the utility of the tools presented so far. At the time of this writing, the 21st National Assembly is still in session, so our focus will be limited to the first 40 months of the assembly, from Jun.\ 2020 to Sep.\ 2023. In South Korea, a bill (legislative proposal) must be supported by at least 10 assembly members. We call them \emph{cosponsors} of the bill. In this study, we do not distinguish between a member who presents the bill (as a representative) and those who cosponsor the bill. For simplicity, the term \emph{cosponsor} is generically used in this section to refer to all members involved in the proposal. Therefore, the number of cosponsored bills by a member denotes the number of bills to which the person has agreed. Similarly, the number of cosponsored bills between two members indicates the number of bills the two members jointly supported. 

A commonly used approach in the social sciences to identify assembly members’ social relationships is to build a network of them using bill cosponsorship information \citep{fowler2006connecting}
In this study, we constructed a graph of 317 assembly members, each representing a single vertex. An edge is established between two members if they have cosponsored at least one bill together. The weight of this edge is defined as the reciprocal of the number of cosponsored bills between two members, so the higher the number of cosponsored bills between two members, the shorter the path length between the corresponding vertices representing these members. A graph is formed by utilizing all 25164 bills proposed over 40 months, provided by the Bill Information System of the National Assembly of South Korea (\url{https://likms.assembly.go.kr/bill/main.do}). The resulting graph is undirected and connected, with 317 vertices and 47657 edges, each edge having a weight. The multiplicities of all vertices are set to 1. We refer to this network as the \emph{assembly network}  for the remainder of this paper. 

As of Sep.\ 30th, 2023, the party composition of the 317 members is as follows (for simplicity, independent members are categorized by their former political party): There are two major parties, the \emph{Democratic Party} with 184 members and the \emph{People Power Party} with 123 members. Additionally, one minor party, the \emph{Justice Party}, has six members. There are also four small parties with only one member each: the \emph{Basic Income Party}, \emph{Hope of Korea}, \emph{Progressive Party}, and \emph{Transition Korea}. Moreover, the 317 vertices comprise two groups: those who served as 21st National Assembly members for 40 months (281) and those who did not (36). The latter group includes members who started their term through a by-election, resigned, or lost their seat for any reason during the 40 months. 

\begin{figure}
\center
\includegraphics[width=\textwidth]{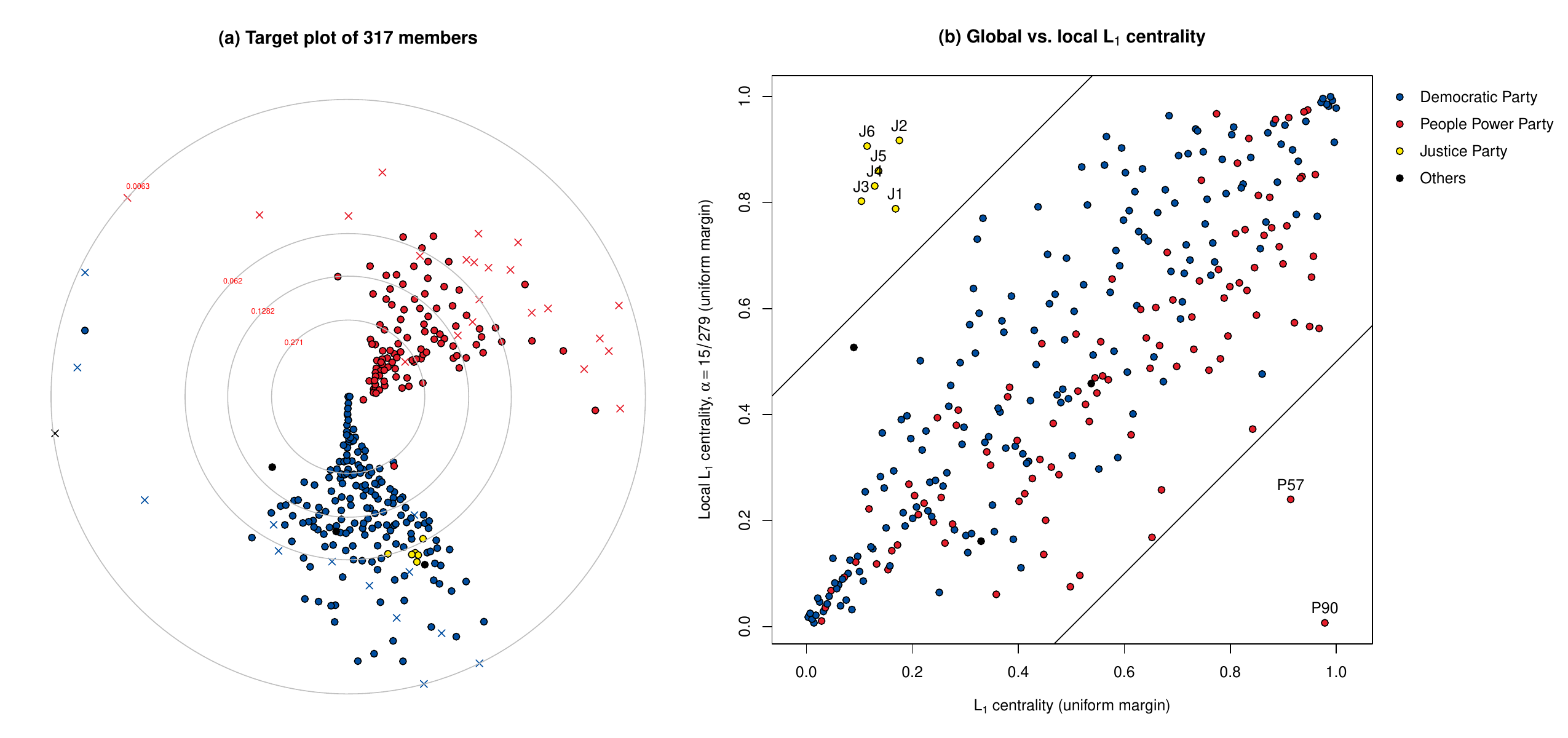}
\vspace{-8mm}
\caption{(a) Target plot of the assembly network with 317 vertices (members). Members who were in the office for the entire 40-month period are denoted by $\circ$, and those who were not are denoted by $\times$. (b) Global versus local $L_1$ centrality with $\alpha = 15/279$. Vertices with an absolute difference of global and local $L_1$ centralities greater than 0.5 in the uniform margin (vertices outside the two diagonals) are indicated with their pseudonyms.} \label{fig:rokassembly-global_vs_local}
\end{figure}

Figure \ref{fig:rokassembly-global_vs_local} (a) is the target plot of the assembly network, which shows that the assembly network is organized into communities by the two major political parties. Furthermore, individuals absent from the office for 40 months exhibit lower global $L_1$ centrality, likely because their time in the assembly is shorter than that of the others, resulting in fewer opportunities to cosponsor bills and build relationships with other members. Therefore, we eliminate these vertices for further analysis, which can be seen as analogous to  \emph{trimming} in traditional statistical analysis. Furthermore, we have removed two chairpersons of the 21st National Assembly with inactive involvement in cosponsoring legislation. As a result, the assembly network is reduced to a subgraph of $n=279$ vertices and 38222 edges. This reduced network will be utilized for the remaining study. 

Next, we performed a multiscale analysis of the reduced assembly network. We calculated the global and local $L_1$ centrality for each member by setting the locality parameter to $\alpha = 15/279$. Figure \ref{fig:rokassembly-global_vs_local} (b) shows a graph comparing the global and local $L_1$ centrality values transformed to have a uniform margin. To facilitate explanation, we assign a random pseudonym to each member with a prefix indicating its political party. A notable observation is that vertices with high $L_1$ centralities also tend to have high local $L_1$ centralities. Similarly, vertices with low $L_1$ centralities also exhibit this phenomenon. Nevertheless, a small number of vertices do not follow this pattern. The six points in the top-left of the plot (J1--J6) and the two vertices in the bottom-right of the plot (P90 and P57) exhibit contrary behavior at the global and local levels. The paradoxical behavior of these eight vertices, which can be compared to the famous Simpson's paradox, is worth examining.

\begin{itemize}
\item All six members of the Justice Party (J1--J6) have low global $L_1$ centrality while demonstrating high local $L_1$ centrality at $\alpha = 15/279$. This can be attributed to the fact that Justice Party members form a tight community with each other: the edges connecting Justice Party members are significantly stronger (i.e., lower edge weights) than those connecting a Justice Party member and a member from another party.

While examining the edges connected to one end of a particular Justice Party member, we consistently observed that the top five edges with the lowest weight were always connected to the other five Justice Party members. Table \ref{tab:jp1} also lists the number of cosponsored bills between the Justice Party members (i.e., the reciprocal of the edge weight connecting two Justice Party members). The magnitude of these values is relatively high for the entire population. The minimum number of cosponsored bills among the members of the Justice Party is 416 bills between J3 and J4, which falls in the 99.64\% quantile in the distribution of the number of cosponsored bills among each member in the reduced assembly network. Hence, it is evident that the edge among the members of the Justice Party is solid.  

Even so, the Justice Party members do not maintain a strong link with other party members. The highest number of cosponsored bills by each Justice Party member with a member of another party ranges between 221 and 317, much lower than the values in Table \ref{tab:jp1}. This implies that the Justice Party is a cohesive community in which bill cosponsorship occurs primarily within the party. Therefore, inside the 15/279 $L_1$ centrality-based neighborhood of each Justice Party member, it is observed that there are four to six other Justice Party members included. As a result, each member of the Justice Party has a high level of centrality within the conditioned graph due to its cohesiveness. 

\begin{table}
\caption{Number of cosponsored bills between each member of the Justice Party, and the total number of cosponsored bills by each member.} \label{tab:jp1} 
\center
\small
\begin{tabular}{r|cccccc}
\hline
  & J1 & J2 & J3 & J4 & J5 & J6\\
\hline
J1 & --- & 439 & 438 & 425 & 438 & 452\\
J2 & 439 & --- & 421 & 418 & 416 & 417\\
J3 & 438 & 421 & --- & 416 & 436 & 430\\
J4 & 425 & 418 & 416 & --- & 431 & 427\\
J5 & 438 & 416 & 436 & 431 & --- & 431\\
J6 & 452 & 417 & 430 & 427 & 431 & ---\\
\hline\hline
Total cosponsored bills & 818 & 566 & 619 & 638 & 660 & 683 \\
\hline
\end{tabular}
\end{table}

However, the total number of bills cosponsored by each member of the Justice Party is too small (last row of Table \ref{tab:jp1}). The median number of cosponsored bills for all network members is 929, and the first quartile is 612. So, while these members are central within the party and their neighbors, the overall strength of those edges is not impressive. As a result, the global $L_1$ centrality is relatively low. This may be the behavior and strategy of a `niche' party that is distinct from the legislative activity patterns of the major parties. They work together effectively, but their small party size limits their ability to engage actively in legislative action. 

\item P57 and P90 exhibit contrasting behaviors to the members of the Justice Party. Their local $L_1$ centrality is low, but their global $L_1$ centrality is high, falling in the 91.40\% (P57) and 97.85\% (P90) quantiles of the global $L_1$ centrality distribution. This is because these two vertices essentially play an important role between the two major parties. As shown from the target plot in Figure \ref{fig:rokassembly-global_vs_local} (a), each of the two major parties establishes a distinct community within the network. A few members, including P57 and P90, serve as a `bridge' vertex between these two communities. 

We started by counting cosponsored bills between and within both major political parties. The distribution is shown in Figure \ref{fig:connection-boxplot}. There is a significant scarcity of bills cosponsored amongst members between the two parties compared to the bills cosponsored by the members of each party. That is, cosponsors of a bill typically consist of only Democratic Party members or only People Power Party members. 

\begin{figure}
\center
\includegraphics[width=0.6\textwidth]{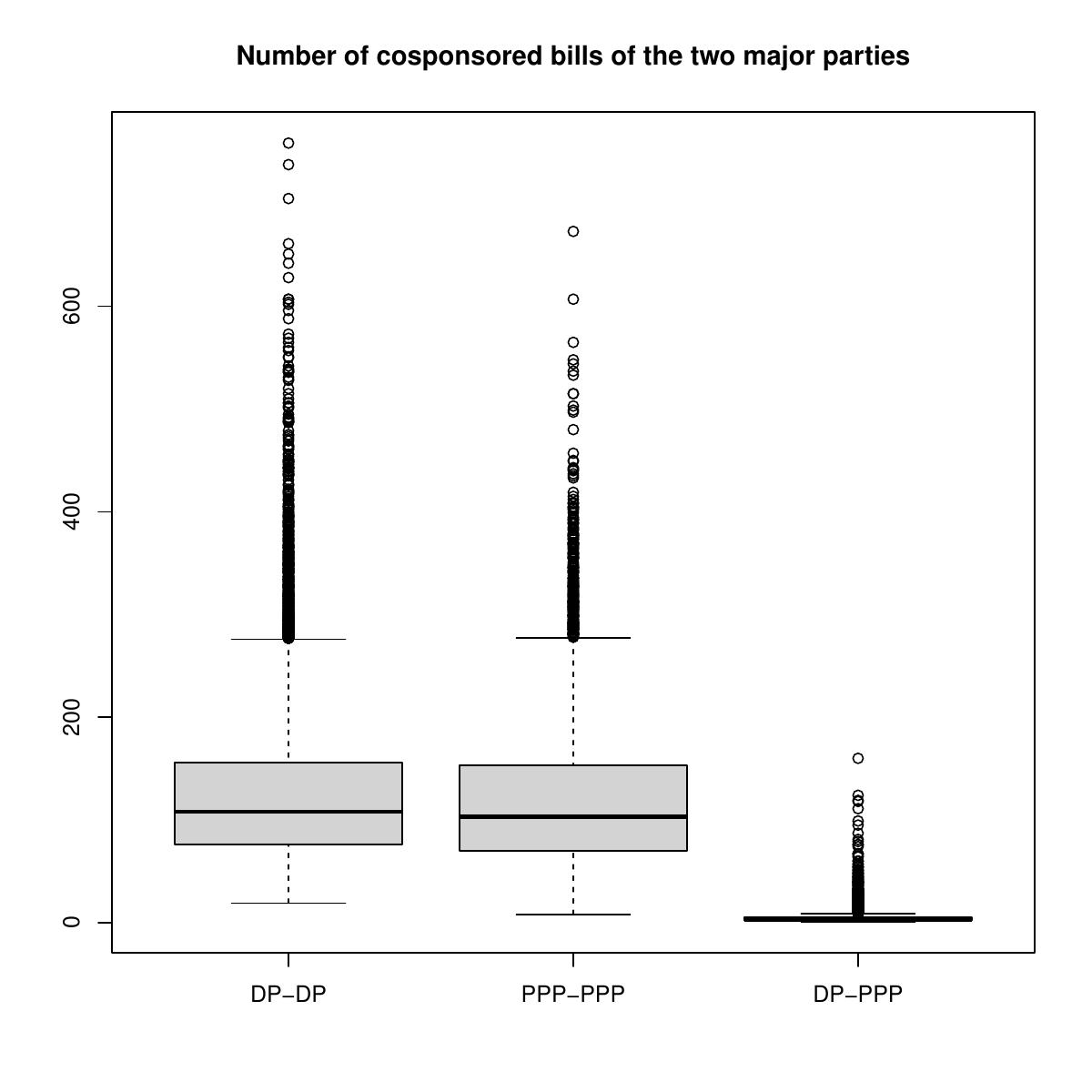}
\vspace{-5mm}
\caption{Distribution of the number of cosponsored bills within the Democratic Party (DP--DP), within the People Power Party (PPP--PPP), and between the two parties (DP--PPP).}\label{fig:connection-boxplot}
\end{figure}

It is important to note that one network member has switched its affiliation from the Democratic Party to the People Power Party, hence being classified as a member of the People Power Party. This individual has cosponsored more bills with members of the Democratic Party than members of the People Power Party. So, except for this member, the top five highest number of cosponsored bills that occurred between the two major parties are as follows: P90--D139 (118 bills), P90--D63 (87 bills), P90--D128 (81 bills), P57--D140 (64 bills), and P8--D140 (60 bills). It is evident that P90 and P57 act as a `bridge' between the two parties. Therefore, P90 and P57, along with vertices D139, D63, D128, D140, and P8, exhibit high global $L_1$ centrality. D128 is in the 72.40\% quantile of the global $L_1$ centrality distribution, and the others are all above the 90\% quantile.

However, among all the vertices with global $L_1$ centrality over the 90\% quantile, P57 has the fewest cosponsored bills (467), and P90 has the second-fewest cosponsored bills (905). Furthermore, among the vertices with global centrality exceeding the 95\% quantile, P90 has the fewest cosponsored bills, which means that the connection between each of the two vertices to its nearby vertices is weak (high edge weight). Therefore, rather than their closest neighbors, vertices with high global $L_1$ centrality are incorporated into the set of $L_1$ centrality-based neighbors for these vertices. This is a phenomenon contrary to that of the Justice Party members. Hence, a small $L_1$ centrality is obtained in the conditioned graph. In summary, the two exceptional vertices, P90 and P57, attain their high global centrality by serving as a `bridge' between the two large communities in the network but have a weak connection to either of the communities. 
\end{itemize}

The analysis presented demonstrates the utility of examining a single graph at multiple locality levels. Indeed, vertices with similar global $L_1$ centrality may exhibit contrasting behavior at a local level and \emph{vice versa}. The proposed global and local $L_1$ centrality captures this specific aspect that cannot be accounted for by other centrality measures. 

\begin{figure}
\center
\includegraphics[width=0.6\textwidth]{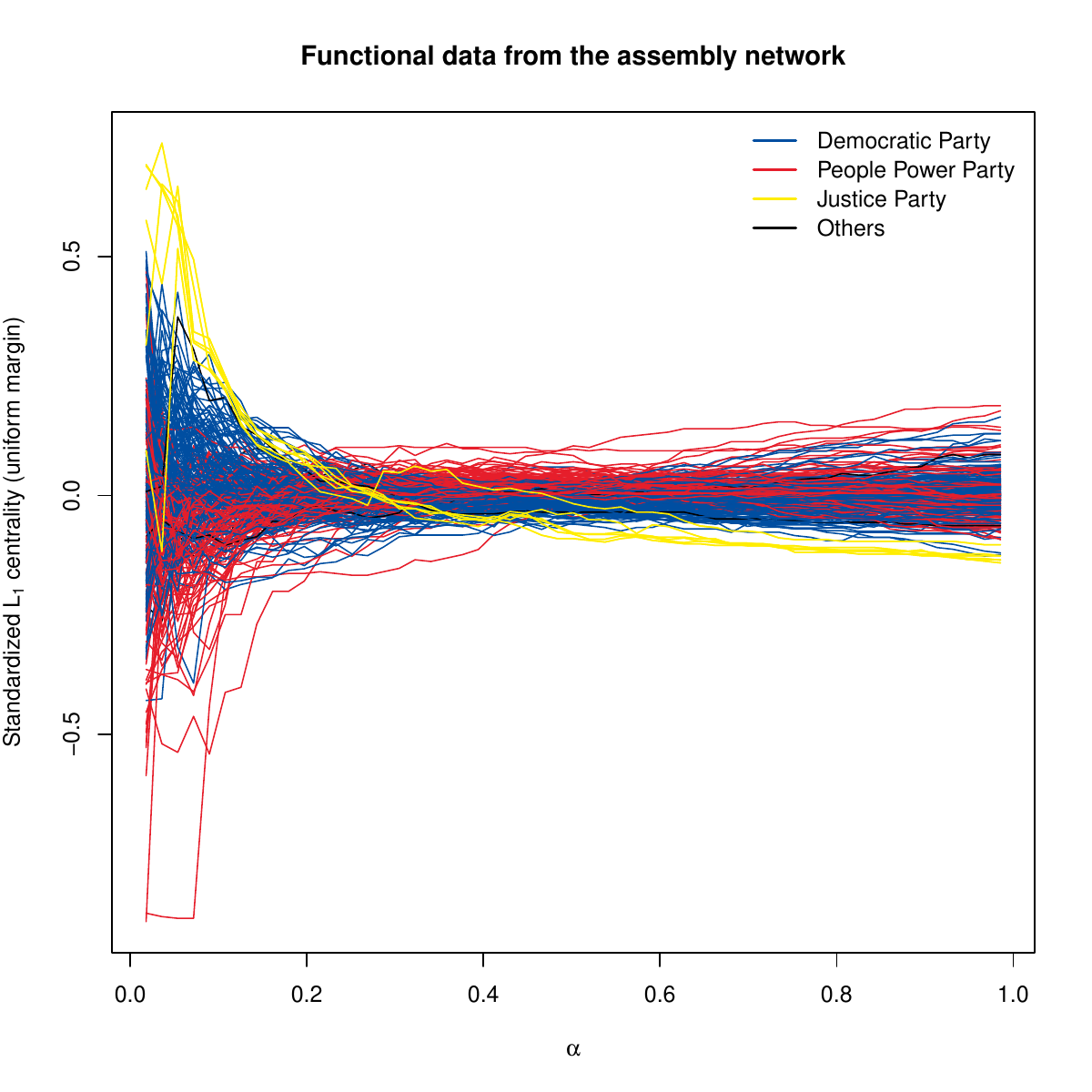}
\vspace{-1em}
\caption{Functional data extracted from the reduced assembly network. Each line represents a member of the reduced assembly network, and it is standardized to have a zero mean.}\label{fig:assembly-fda}
\end{figure}

In this analysis, we used two values of $\alpha=15/279$ and 1. However, a more refined multiscale analysis can be performed with densely chosen values of $\alpha$. By adjusting the locality level, we may create a functional data set in which each function corresponds to a member of the reduced assembly network. Figure \ref{fig:assembly-fda} shows the variation in the local $L_1$ centrality of each vertex over $\alpha=5/279, 10/279, \dots, 275/279$. The local $L_1$ centrality values are transformed to a uniform margin for each locality level. For easier comprehension, each curve is then standardized to have a mean of zero, meaning that the average of the functional values is zero for each curve. The curves representing the Justice Party members deviate from the other curves. Collecting structural information of each vertex as a function can be more informative than other single-valued centrality measures. This functional data can be utilized for various analyses. For example, one can classify or cluster vertices or understand the relationship between the structural attribute of each vertex and other relevant characteristics. Extensive research on functional data analysis can be utilized with this data set \citep[see, e.g.,][]{ramsay2005fda}. 

\begin{figure}
\center
\includegraphics[width=\textwidth]{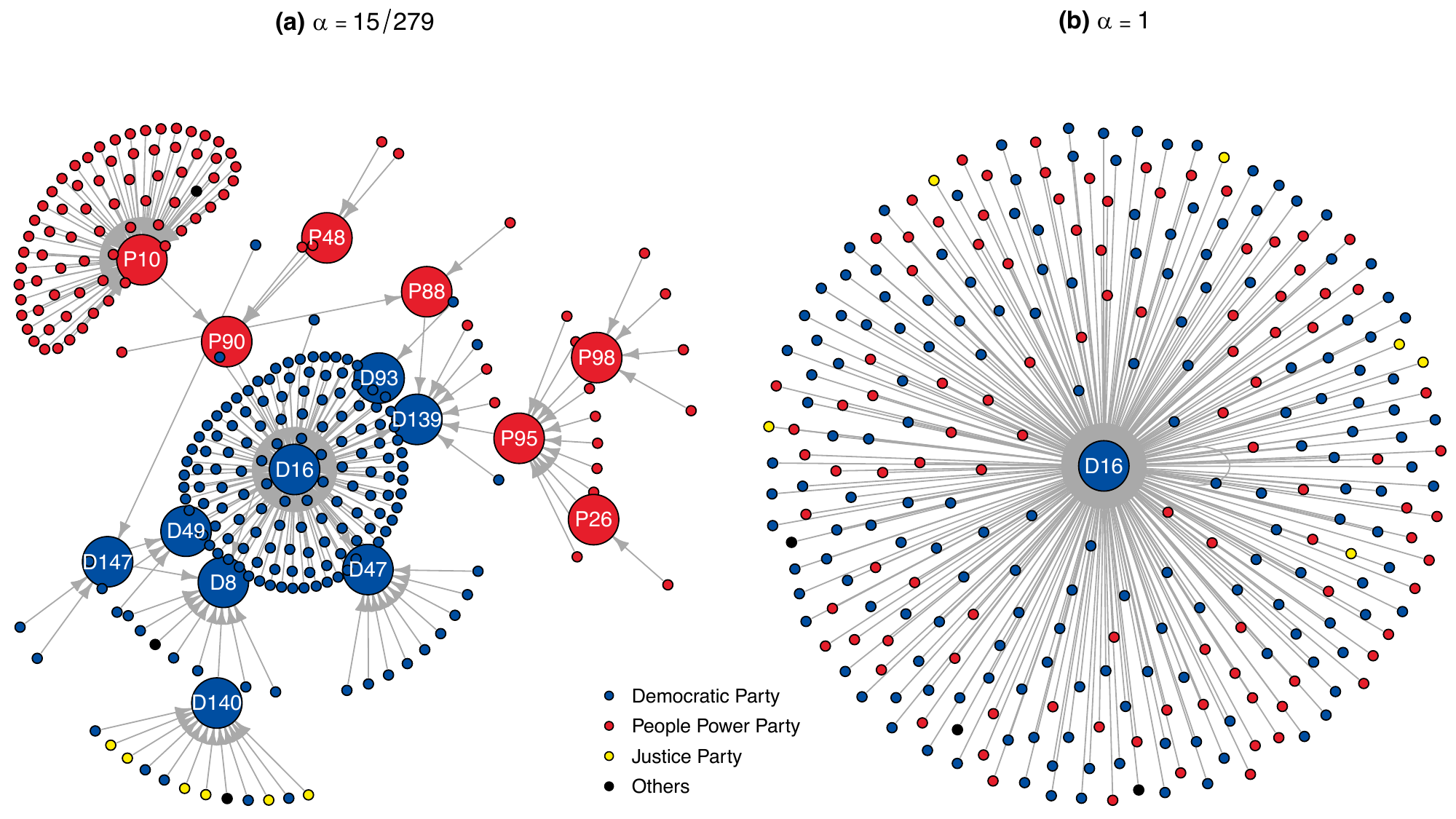}
\vspace{-7mm}
\caption{Multiscale edge representation of the reduced assembly network with two locality levels: $\alpha = 15/279$ and $\alpha = 1$. Larger vertices with pseudonyms labeled represent the local medians.}\label{fig:assembly-edge}
\end{figure}

Finally, we present a multiscale edge representation of the reduced assembly network with two locality levels, $\alpha = 15/279$ and $1$. In Figure \ref{fig:assembly-edge}, the local medians are represented by larger vertices, each labeled with the pseudonym of its member. Panel (a) reveals the presence of 15 local medians, with D16 being identified as the local median originating from nearly half of the vertices. As shown in panel (b), D16 serves as the graph median of the reduced assembly network. It would be worthwhile to investigate some local medians, such as D139, which serves as the local median for a number of People Power Party members, and D140, which is designated as the local median for all six Justice Party members. Altering the value of $\alpha$ to different values makes a comprehensive multiscale representation of the assembly network possible. Conversely, a plot with all 38222 edges would be incomprehensible.  

\section{Discussion} \label{sec:discussion}

In this paper, we have introduced the $L_1$ centrality for evaluating vertex centrality in graphs and related tools for multiscale and graphical analysis. The proposed method borrows ideas from existing literature on the data depth for multivariate data and particular instruments and methods used in the field. The tools provided in this paper are only a first step. There are many potential applications beyond the current discussion. 
The literature on data depth suggests various areas where the $L_1$ centrality can be used successfully. A few examples from the data depth literature include classification and clustering \citep{jornsten2004clustering,li2012dd,paindaveine2013depth}, missing value imputation \citep{mozharovskyi2019nonparametric}, and nonparametric test \citep{liu1993quality}. 

This study only considered undirected graphs to define measures and tools. We briefly explore the potential extension of the concepts to directed graphs. In the literature, the concept of ‘centralness’ in directed graphs is divided into two distinct notions: prestige and centrality \citep[Chapter 5.3]{wasserman_faust_1994}. The prestige measure estimates the level of importance of a vertex in terms of \emph{receiving} a choice. Conversely, the centrality measure in directed graphs assesses the level of importance in terms of \emph{giving} a choice. 
It is feasible to extend the $L_1$ centrality to encompass directed graphs. Define the $L_1$ prestige for directed graphs in the following way. The graph median for prestige measure is redefined as the vertex $v_i$ that minimizes $\sum_{j=1}^n\eta_jd(v_j,v_i)$, where $d(v_j,v_i)$ denotes the geodesic distance from $v_j$ to $v_i$. Here, $d(v_i,v_j)$ is not necessarily equal to $d(v_j,v_i)$, which means it is not a distance function. Assuming that every vertex can be reached from another, the $L_1$ prestige can be defined according to equation \eqref{eq:l1-cent} with the indices in the function $d$ swapped. Nevertheless, since the triangle inequality, such as $d(v_i,v_k)-d(v_i,v_j)\leq d(v_k,v_j)$, does not necessarily hold---since $d$ is not a distance function---the resulting prestige measure cannot be ensured to fall within the range of $[0,1]$. Together with the analogous theorem and propositions in this paper that apply to directed graphs, we plan to carry out a separate study on this topic elsewhere. 


Next, recall that the concept of $L_1$ centrality is derived from the $L_1$ data depth. It is of interest to explore the theoretical connection between these two notions. A possible link can be made in the latent space model for graph data \citep{smith2019geometry}. In this model, edges are assumed to be generated depending on the positions of points in the underlying geometric space that represent the two vertices. With this model, we may ask if the $L_1$ centrality is a `good' estimate of the $L_1$ depth w.r.t.\ the underlying probability distribution in the latent space. Then, we can consider the $L_1$ centrality as a desirable \emph{statistic} rather than an ad-hoc \emph{measure}. This can also construct a comprehensive distribution theory for a centrality statistic. For example, the large sample ($n\to\infty$) behaviors of the centralities. We have established a proposition that may be useful for further study.
\begin{prop}\label{prop:as-conv}
Let $M\subset \RR^d$ be a convex $d$-dimensional space. Given $X_1=x_1\in M$, suppose that $X_2, \dots, X_n$ are i.i.d.\ samples from some probability distribution $P$ supported on $M$, which we assume to have a density $p$, bounded away from zero: $p(z) \geq c_p > 0$ for all $z\in M$. Multiplicities of $x_1,X_2,\dots,X_n$ are $\eta_1,\eta_2,\dots,\eta_n$ respectively, and $\etadot = \sum_{i=1}^n\eta_i$. Then, the following strong consistency to the $L_1$ depth of $x_1$ w.r.t.\ $P$, denoted as $D(x_1)$, holds. That is, with the usual Euclidean norm $\|\cdot\|$,
\begin{align*}
1-\max_{j=2,\dots,n}\left\{\frac{\sum_{i=1}^n\eta_i(\|x_1-X_i\|-\|X_j-X_i\|)}{\etadot \|X_j-x_1\|}\right\}^+ \xrightarrow{n\to\infty} D(x_1) \quad\text{a.s.}
\end{align*}
\end{prop}
The LHS is similar to equation \eqref{eq:l1-cent}, except that instead of using geodesic distance in the graph, it uses the Euclidean norm. Thus, if the geodesic distance in the graph approaches the Euclidean distance in the underlying geometric space at a suitable speed, it is possible to establish the convergence of the $L_1$ centrality to the $L_1$ depth in the latent space. Much literature proves that the geodesic distance in the graph converges to the geodesic distance in various latent space models \citep{alamgir2012shortest,10.1214/15-AAP1162}. Nevertheless, there is uncertainty about which latent space models may effectively achieve the desired convergence speed and the conditions necessary to establish a concrete connection between the two concepts. This topic can be explored in future research, perhaps with a graph centrality based on the notion of another data depth function.

\section*{Acknowledgements}

This research was supported by the National Research Foundation of Korea (NRF) funded by the Korea government (2021R1A2C1091357). 

\section*{Data Availability}
The MCU movie network and the assembly network are available from the R package \textbf{L1centrality} (\url{https://CRAN.R-project.org/package=L1centrality}).

\bibliographystyle{abbrvnat}
{\small\bibliography{bibs}}

\newpage
\begin{appendix}
\section*{Appendix}

\section{Proofs}

\subsection{Proof of Theorem \ref{thm:property}} \label{app:property-proof}

\begin{proof}
\begin{itemize}
\item[(P1)] This is immediate from equation \eqref{eq:l1-cent}.
\item[(P2)] It suffices to show that $v_i$ is the graph median if $\eta_i/\etadot \geq 1/2$ and the unique graph median if $\eta_i/\etadot > 1/2$. Observe that for $i'\neq i$, $\sum_{j=1}^n \eta_j \{d(v_i, v_j) - d(v_{i'},v_j)\} \leq \sum_{j\neq i,i'} \eta_j d(v_i, v_{i'}) + \eta_{i'}d(v_i,v_{i'}) - \eta_id(v_{i'},v_i) = (\etadot - 2\eta_i)d(v_{i'},v_i)$ by the triangle inequality. This yields that $C(v_i) \geq 1- \{(\etadot - 2\eta_i)/\etadot\}^+ = 1$ if $\eta_i/\etadot \geq 1/2$. In addition, $C(v_{i'})\leq 1 - \{(2\eta_i - \etadot)/\etadot\}^+ < 1$ if $\eta_i/\etadot > 1/2$.
\item[(P3)] Suppose that $v_i$ is not a graph median. Due to (P2), $v_i$ becomes the graph median when multiplicity is incremented by $1-2\eta_i/\etadot$. This is because the multiplicity of $v_i$ is $1-\eta_i/\etadot$ after incrementing, and the sum of all vertices' multiplicities is $2-2\eta_i/\etadot$. Hence, $1-C(v_i)\leq 1-2\eta_i/\etadot$. If $v_i$ is the graph median, $C(v_i) = 1$. Thus, $C(v_i)\geq \min\{2\eta_i/\etadot, 1\}$.
\item[(P4)] Since the subgraph induced by deleting vertex $v_1$ is connected, it means that $\sum_{i\neq 1}\eta_i d(v_i,v_j)$ converges to a finite value for any $j\neq 1$, as $v_1$ is moved to infinity. Observe that
\begin{align*}
\max_{j\neq 1}\frac{\sum_{i=1}^n \eta_i(d(v_i,v_1)-d(v_i,v_j))}{\etadot d(v_1,v_j)} 
&=\max_{j\neq 1}\frac{\sum_{i\neq 1} \eta_id(v_i,v_1)-\sum_{i\neq 1} \eta_i d(v_i,v_j) - \eta_1 d(v_1,v_j)}{\etadot d(v_1,v_j)} \\
&\to \frac{(\etadot - \eta_1) + 0 - \eta_1}{\etadot}.
\end{align*}
Hence, $C(v_1) \to 1 - \{(\etadot - 2\eta_1)/\etadot\}^+ = \min\{2\eta_1/\etadot, 1\}$. \qedhere
\end{itemize}
\end{proof}

\subsection{Proof of Proposition \ref{prop:d-symm}}

\begin{proof}
Since $L_1$ centrality of $v_i$ in the symmetrized graph w.r.t.\ $v_i$ is equivalent to the $L_1$ centrality of $v_i$ in the original graph with multiplicity $\eta_i$ replaced to $\etadot + \eta_i$ (Proposition \ref{prop:new-distance-mat}), Theorem \ref{thm:property} (P2) implies that $v_i$ is the  graph median of the symmetrized graph since $(\etadot + \eta_i)/(2\etadot) \geq 1/2$. If $\eta_i >0$, $(\etadot + \eta_i)/(2\etadot) > 1/2$, which yields that $v_i$ is the unique graph median.
\end{proof}

\subsection{Proof of Proposition \ref{prop:new-distance-mat}}

\begin{proof}
Without loss of generality, suppose that $G$ is symmetrized w.r.t.\ vertex $v_1$. Denote the copied vertices with `prime' on their index, e.g., $v_{k'}$ is the copy of $v_k$. In the symmetrized graph, the $L_1$ centrality of vertex $v_k$ is
\begin{align*}
1 - \max_{j \neq k}\left\{\frac{\text{num}}{2\etadot d(v_j,v_k)}\right\}^+.
\end{align*}
Here, the numerator is given as
\begin{align*}
\text{num} =& \sum_{i=1}^n \eta_i d(v_i,v_k) + \sum_{i=1}^n \eta_i (d(v_i,v_1)+d(v_1,v_k))  - \sum_{i=1}^n \eta_i d(v_i,v_j) - \sum_{i=1}^n \eta_i (d(v_i,v_1)+d(v_1,v_j)) \\
=& \sum_{i=1}^n \eta_i \{d(v_i,v_k)-d(v_i,v_j)\} + \etadot (d(v_1,v_k)-d(v_1,v_j)) \\
=& \sum_{i=1}^n \eta_i' (d(v_i,v_k)-d(v_i,v_j)), 
\end{align*}
where 
\[
\eta_i' = \begin{cases}
\eta_i, &(i\neq 1)\\
\etadot + \eta_1, &(i=1)
\end{cases}.
\]
Hence, the $L_1$ centrality of vertex $v_k$ in the symmetrized graph is equivalent to the $L_1$ centrality of $v_k$ in the original graph with multiplicities $\bfeta = (\etadot + \eta_1, \eta_2,\dots,\eta_n)^\top$.
\end{proof}

\subsection{Proof of Proposition \ref{prop:as-conv}}

\begin{proof}
Let $\bar\be(x_1) \coloneqq \int_{M\setminus\{x_1\}} \frac{y-x_1}{\|y-x_1\|}\dif P(y)$ and $\bar\be_n(x_1)\coloneqq\sum_{i=2}^n \frac{\eta_i}{\etadot}\frac{x_i-x_1}{\|x_i-x_1\|}$. According to \cite{vardi2000multivariate} and \cite{tian2002l1depth}, $\|\bar\be(x_1)\| = 1-D(x_1)$, and $\bar\be_n(x_1)\to\bar\be(x_1)$ a.s.\ as $n\to\infty$ by the strong law of large numbers (SLLN). 

If $\|\bar\be(x_1)\| = 0$, \cite{vardi2000multivariate} proved that $x_1$ is the $L_1$ median w.r.t.\ $P$. Hence, by the SLLN, the numerator of the LHS in the proposition converges to a nonpositive number a.s., which means that the LHS converges to $D(x_1) = 1$ a.s. In the rest of the proof, we assume $\|\bar\be(x_1)\| > 0$. 

Set $X^* = x_1 + h_n \bar\be(x_1)$ and let $X^\dagger$ be the closest random point to $X^*$ among $X_2,\dots,X_n$. Here, $h_n\downarrow 0$ and $nh_n^d - \log n \to \infty$, as $n\to\infty$ (e.g., take $h_n = (\log n)^{-1/d}$). Then, for a sufficiently large $n$,
\begin{align*}
\Pr(\|X^\dagger-X^*\| \geq h_n\epsilon) \leq (1-c_d(h_n\epsilon)^d)^{n-1}
\leq \exp\{-(n-1)c_d(h_n\epsilon)^d\},
\end{align*}
for some constant $c_d > 0$ that depends on $d$. Elaborating, due to the convexity of $M$, a ball with center $X^*$ and radius $h_n\epsilon$ will be inside $M$, for a sufficiently large $n$. The first inequality is derived from the probability that all $X_2,\dots,X_n$ to be located outside this ball, and the second inequality is immediate from $1-x\leq e^{-x}$, which holds for all $x\in\RR$.

Since $nh_n^d - \log n \to \infty$ as $n\to\infty$, Borel--Cantelli lemma can be applied to say that $\|X^\dagger-X^*\|/h_n \to 0$ a.s.\ as $n\to\infty$. This further implies that
\begin{align*}
\frac{1}{h_n}\sup_{i=1,\dots,n}\left|\|X^\dagger- X_i\| - \|X^*-X_i\|\right| \leq \frac{1}{h_n} \|X^\dagger-X^*\| \to 0\quad\text{a.s.}
\end{align*}
and thus, with probability one,
\begin{align*}
\|X^\dagger- x_1\| = \|X^*- x_1\| + o(h_n) = \|X^*- x_1\|\left(1 + \frac{o(h_n)}{h_n\|\bar\be(x_1)\|}\right) = \|X^*-x_1\|[1+o(1)],
\end{align*}
where $o(a_n)/a_n\to 0$ as $n\to\infty$.

Using these facts, we get that, with probability one,
\begin{align}
\max_{j=2,\dots,n}\left\{\frac{\sum_{i=1}^n\eta_i(\|x_1-X_i\|-\|X_j-X_i\|)}{\etadot \|X_j-x_1\|}\right\}^+ 
&\geq \left\{\frac{\sum_{i=1}^n\eta_i(\|x_1-X_i\|-\|X^\dagger-X_i\|)}{\etadot \|X^\dagger-x_1\|}\right\}^+ \nonumber\\
&= \left\{\frac{\sum_{i=1}^n\eta_i(\|x_1-X_i\|-\|X^*-X_i\|)}{\etadot \|X^*-x_1\|}\right\}^+  + o(1). \label{eq:a1}
\end{align}

By the dominated convergence theorem, the RHS of equation \eqref{eq:a1} converges to the following value a.s.
\begin{align*}
\left\{\int_{M\setminus\{x_1\}}\lim_{n\to\infty}\frac{\|x_1-y\|-\|x_1+h_n\bar\be(x_1) - y\|}{\|h_n\bar\be(x_1)\|}\dif P(y) \right\}^+
&= \|\bar\be(x_1)\| = 1-D(x_1),
\end{align*}
where the first equality is derived by differentiation.

However, observe that
\begin{align}
\max_{j=2,\dots,n}\left\{\frac{\sum_{i=1}^n\eta_i(\|x_1-X_i\|-\|X_j-X_i\|)}{\etadot \|X_j-x_1\|}\right\}^+ 
&\leq \sup_{y\in M\setminus\{x_1\}}\left\{\frac{\sum_{i=1}^n\eta_i(\|x_1-X_i\|-\|y-X_i\|)}{\etadot \|y-x_1\|}\right\}^+. \label{eq:a2}
\end{align}
The RHS of equation \eqref{eq:a2} is equivalent to $(\|\bar\be_n(x_1)\|-\eta_1/\etadot)^+$ by \cite{vardi2000multivariate}, and hence, also converges to $1-D(x_1)$ a.s., and the proof is complete.
\end{proof}

\section{Comparing $L_1$ Centrality Measure to Other Measures} \label{app:compare}

\begin{figure}
\center
\includegraphics[width=\textwidth]{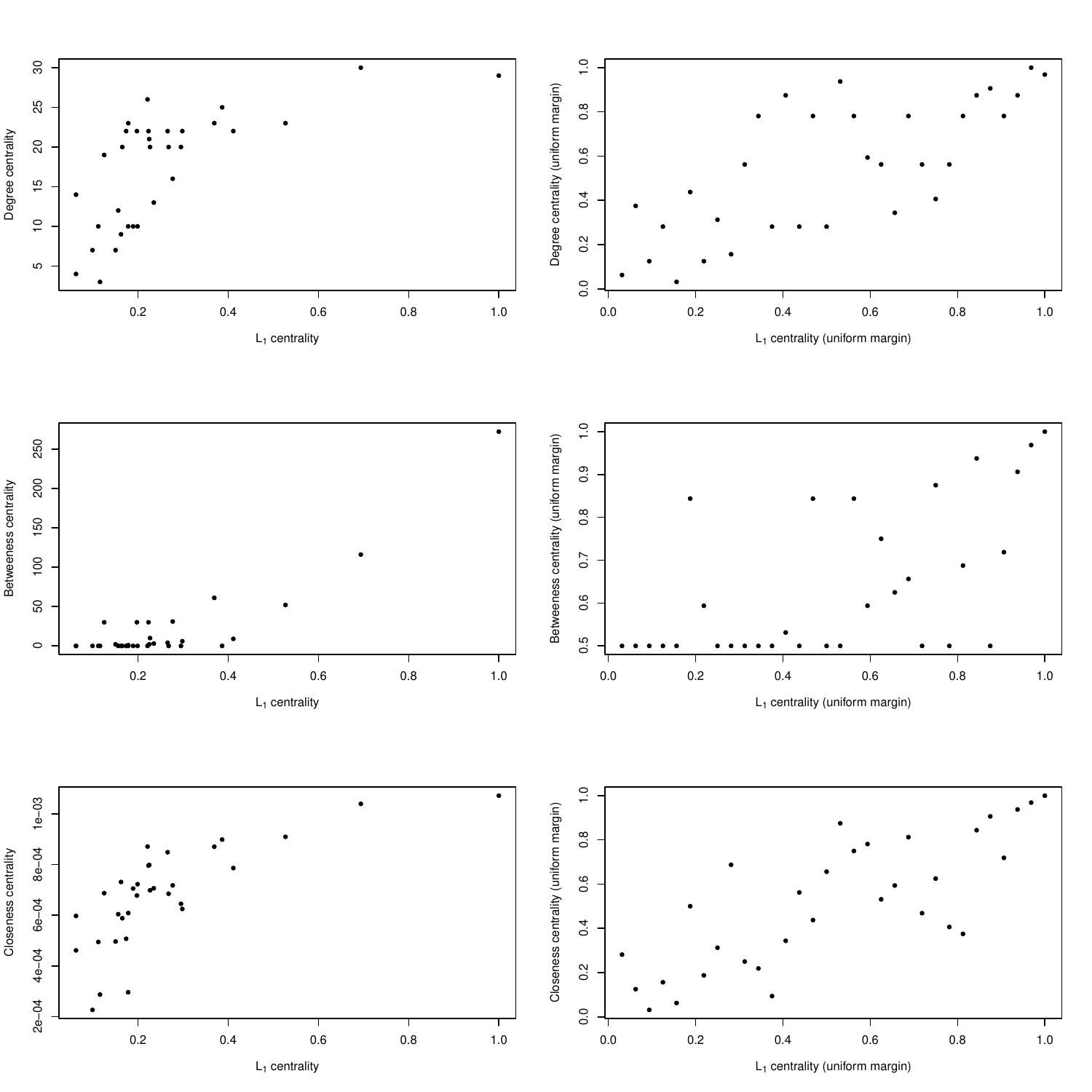}
\vspace{-7mm}
\caption{Degree centrality, betweenness centrality, and closeness centrality applied to the MCU movie network. (Left) Original centrality values. (Right) Centralities transformed to a uniform margin.}\label{fig:mcu-comparison}
\end{figure}

The first column of Figure \ref{fig:mcu-comparison} shows the $L_1$ centrality versus the three centralities. The second column transforms computed centralities to a uniform margin for easy comparison, i.e., the lowest centrality is converted to $1/n$, the second lowest to $2/n$, and so on. Thus, if the rankings from the centrality measures are the same, every point would fall on the $45^{\circ}$ diagonal line, which is not valid for all plots in the second column.

\end{appendix}

\end{document}